\documentclass[fleqn,usenatbib]{mnras}

\usepackage{newtxtext,newtxmath}
\usepackage{mathrsfs}

\usepackage[T1]{fontenc}

\usepackage{graphicx}	
\usepackage{amsmath}	
\usepackage{soul}
\usepackage{xcolor}

\title[White-dwarf relativistic pulsations]{General relativistic pulsations of ultra-massive ZZ Ceti stars}

\author[C\'orsico et al.]{
Alejandro H. C\'orsico$^{1,2}$\thanks{E-mail: acorsico@fcaglp.unlp.edu.ar},
S. Reece Boston$^{3,4}$,
Leandro G. Althaus$^{1,2}$,
Mukremin Kilic$^{5}$,
\newauthor 
S. O. Kepler$^{6}$,
Mar\'ia E. Camisassa$^{7}$ and
Santiago Torres$^{7,8}$,
\\
$^{1}$     Facultad de Ciencias Astron\'omicas y Geof\'{\i}sicas, 
           Universidad Nacional de La Plata, 
           Paseo del Bosque s/n, 1900 
           La Plata, 
           Argentina
           \\
$^{2}$ Instituto de Astrof\'isica de La Plata, UNLP-CONICET,
           Paseo del Bosque s/n, 
           1900 La Plata, 
           Argentina\\
$^{3}$ University of North Carolina at Chapel Hill, Chapel Hill,
           NC 27599, USA\\
$^{4}$ Computer Science and Mathematics Division, Oak Ridge National Laboratory, Oak Ridge, TN 37830, USA\\
$^{5}$ Homer L. Dodge Department of Physics and Astronomy,
           University of Oklahoma, 440 W. Brooks St., Norman,
           OK 73019, USA\\
$^{6}$ Instituto de F\'isica, Universidade Federal do Rio Grande do Sul, 
           91501-900 Porto-Alegre, RS, Brazil\\
$^{7}$ Departament de Física, Universitat Politècnica de Catalunya,
           c/Esteve Terrades 5, 08860 Castelldefels, Spain\\
$^{8}$ Institute for Space Studies of Catalonia, 
           c/Gran Capita 2--4, 
           Edif. Nexus 104, 
           08034 Barcelona, 
           Spain
}         
\date{Accepted XXX. Received YYY; in original form ZZZ}

\pubyear{2023}

\begin{document}
\label{firstpage}
\pagerange{\pageref{firstpage}--\pageref{lastpage}}
\maketitle

\begin{abstract}
Ultra-massive  white dwarf stars are currently being discovered at a considerable rate, thanks to surveys such as the {\it Gaia} space mission. These dense and compact stellar remnants likely play a major role in type Ia supernova explosions.  It is possible to probe the interiors of ultra-massive white dwarfs through asteroseismology. In the case of the most massive white dwarfs, General Relativity could affect their structure and pulsations substantially. In this work, we present results of relativistic pulsation calculations employing relativistic ultra-massive ONe-core white dwarf models with hydrogen-rich atmospheres and masses ranging from $1.29$ to $1.369 M_{\sun}$  with the aim of assessing the impact of General Relativity on the adiabatic gravity ($g$)-mode period spectrum of very-high mass ZZ Ceti stars.  Employing the relativistic Cowling approximation for the pulsation analysis, we find that the critical buoyancy (Brunt-V\"ais\"al\"a) and acoustic (Lamb) frequencies are larger for the relativistic case, compared to the Newtonian case, due to the relativistic white dwarf models having smaller radii and higher gravities for a fixed stellar mass. In addition, the $g$-mode periods are shorter in the relativistic case than in the Newtonian computations, with relative differences of up to $\sim 50$ \% for the highest-mass models ($1.369 M_{\sun}$) and for effective temperatures typical of the ZZ Ceti instability strip. Hence, the effects of General Relativity on the structure, evolution, and pulsations of white dwarfs with masses larger than $\sim 1.29 M_{\odot}$ cannot be ignored in the asteroseismological analysis of ultra-massive ZZ Ceti stars.
\end{abstract}

\begin{keywords}
stars:  evolution  ---  stars: interiors  ---  stars:  white dwarfs --- stars: pulsations --- asteroseismology --- relativistic processes
  
\end{keywords}


\section{Introduction}
\label{introduction}

ZZ Ceti variables are pulsating DA (H-rich atmosphere) white dwarf (WD) stars with effective temperatures in the range
$10\,500 \lesssim T_{\rm eff} \lesssim 13\,500$ K  and surface gravities in the interval  $7.5\lesssim \log g \lesssim 9.35$.
They exhibit periods from $\sim 100$ s to $\sim 1400$ s due to nonradial gravity($g$) modes with harmonic degree
$\ell= 1$ and $\ell= 2$ \citep{2008ARA&A..46..157W, 2008PASP..120.1043F,
2010A&ARv..18..471A}. The interiors of these compact stars, which constitute the evolutionary end of most stars in the Universe, can be investigated through the powerful tool of asteroseismology by comparing the observed periods with 
theoretical periods computed using large grids of WD stellar models  \citep[e.g.][]{2019A&ARv..27....7C}. 

Although most ZZ Ceti stars
have masses between $\sim 0.5$ and $\sim 0.8~M_{\odot}$, at least seven ultra-massive ($M_{\star} \gtrsim 1.05 M_{\odot}$)
ZZ Ceti stars have been discovered so far: BPM~37093 \citep[$M_{\star}=1.13M_{\odot}$;][]{1992ApJ...390L..89K,2017ApJ...848...11B},
GD~518 \citep[$M_{\star}=1.24~M_{\odot}$;][]{2013ApJ...771L...2H},
SDSS~J084021.23+522217.4 \citep[$M_{\star}= 1.16~M_{\odot}$;][]{2017MNRAS.468..239C}, 
WD~J212402.03$-$600100.05 \citep[$M_{\star}= 1.16~M_{\odot}$;][]{2019MNRAS.486.4574R},
J0204+8713 and J0551+4135  \citep[$M_{\star}= 1.05~M_{\odot}$ and    $M_{\star}= 1.13~M_{\odot}$, respectively;][]{2020AJ....160..252V},
and WD~J004917.14$-$252556.81 \citep[$M_{\star}\sim  1.30~M_{\odot}$;][]{2023MNRAS.522.2181K}.
With such a high stellar mass, the latter is the most massive pulsating WD currently known. The discovery and characterisation of pulsating ultra-massive WDs through asteroseismology is important for understanding the supernovae type Ia explosions. We know that accreting CO-core WDs are the progenitors of these explosions \citep[e.g.,][]{nugent11,maoz14}, but we have not been able to probe the interior structure of such WDs near the Chandrasekhar limit.

Modern photometric data of pulsating WDs collected by spacecrafts such as the ongoing Transiting Exoplanet Survey Satellite mission \citep[{\it TESS};][]{2014SPIE.9143E..20R} and the already finished 
{\it Kepler}/{\it K2} space mission \citep{2010Sci...327..977B, 2014PASP..126..398H}, 
brought along revolutionary improvements to the field of WD asteroseismology in at least two aspects \citep{2020FrASS...7...47C, 2022BAAA...63...48C}. First, the space missions provide pulsation periods with an unprecedented precision. Indeed, the observational precision limit of {\it TESS} for the pulsation periods is of order $\sim 10^{-4}$  s or even smaller \citep{2022FrASS...9.9045G}. Second, these space missions also enable the discovery of large numbers of new pulsating WDs. 
For example, \citet{2022MNRAS.511.1574R} used the {\it TESS} data from the first three years of the mission, for Sectors 1 through 39, to
identify 74 new  ZZ Ceti stars, which increased the number of already known 
ZZ Cetis by
$\sim 20$ per cent. It is likely that many more pulsating WDs, not only average-mass ($M_{\star} \sim 0.60~M_{\sun}$) objects, but also ultra-massive WDs, will be identified by {\it TESS} and other future space telescopes such as the Ultraviolet Transient Astronomy Satellite \citep[ULTRASAT,][]{benami22} in the coming years, though {\it TESS}'s relatively small aperture limits its ability to observe intrinsically fainter massive WDs. In addition, large-scale wide-field ground-based photometric surveys like the Vera C. Rubin Observatory's Legacy Survey of Space and Time and the BlackGEM \citep{groot22} will significantly increase the population of WD pulsators, including massive WDs.

The use of space telescopes for WD asteroseismology has opened up a new window into the interiors of these stars and led to some new and interesting questions. For example, the availability of pulsation periods with high precision supplied by modern space-based photometric observations has, for the first time, raised the question of whether it is possible to detect very subtle effects in the observed period patterns, such as the signatures of the current experimental $^{12}$C$(\alpha, \gamma)^{16}$O reaction rate probability distribution function \citep{2022ApJ...935...21C}, or the possible impact of General Relativity (GR) on the pulsation periods of ZZ Ceti stars \citep{2023arXiv230413055R}. In particular, the possibility that relativistic effects can be larger than the uncertainties in the observed periods when measured with space missions has led \cite{2023arXiv230413055R} to conclude that, for average mass WDs, the relative differences between periods in the Newtonian and relativistic calculations can be larger than the observational precision with which the periods are measured. Hence, to fully exploit the unprecedented quality of the observational data from {\it TESS} and similar space missions, it is necessary to take into account the  GR effects on the structure and pulsations of WDs. 

The impact of GR  is stronger as we consider more massive WD configurations. In particular, WDs with masses close to the Chandrasekhar mass ($M_{\rm Ch}\sim 1.4 M_{\sun}$). \citet{2018GReGr..50...38C, 2021ApJ...921..138N} and \citet{2022A&A...668A..58A}
used static WD models and evolutionary ONe-core WD configurations, respectively, to explore the effects of GR  on the structure of ultra-massive WDs. These investigations found that 
GR strongly impacts the radius and surface gravity of ultra-massive WDs. In addition, \cite{2022A&A...668A..58A} found that  GR  leads to important changes in cooling ages and in mass-radius relationships when compared with Newtonian computations. Furthermore, 
\cite{2023MNRAS.523.4492A}  have extended the relativistic computations to CO-core ultra-massive WD models.

In the present work, we aim to assess the impact of GR on the 
$g$-mode period spectra of ultra-massive ZZ Ceti stars with masses $\gtrsim 1.29 M_{\sun}$.  This is the lower limit for the WD mass from which the effects of GR begin to be relevant \citep{2022A&A...668A..58A}. Our analysis is complementary to that of \cite{2023arXiv230413055R}, which is focused on average-mass pulsating DA WDs ($\sim 0.60 M_{\sun}$; the bulk of pulsating WD population). 
For these average-mass DA WDs, the difference of Newtonian physics and GR was shown to be on the order of the surface gravitational redshift $z\sim10^{-4}$, though for stars with very high central concentration of mass this difference could be an order of magnitude larger.  Since the ultra-massive WDs are highly centrally condensed, GR might be even more important for these objects. The study of ultra-massive WDs is of particular interest at present, given the increasing rate of discovery of these objects \citep{2018ApJ...861L..13G, 2020ApJ...898...84K, 2021MNRAS.503.5397K,    2020NatAs...4..663H, 2021Natur.595...39C, 2022MNRAS.511.5462T,  2023MNRAS.518.2341K}
and the prospect of finding pulsating ultra-massive WDs more massive than WD~J004917.14$-$252556.81 \citep{2023MNRAS.522.2181K}. This last point is particularly relevant in view of the  capabilities of the current (e.g. {\it TESS}) and upcoming (e.g., {\it ULTRASAT, LSST, BlackGEM}) surveys.

The formalism of stellar pulsations in GR began with \citet{1967ApJ...149..591T}, using the Regge-Wheeler gauge to treat the pulsations as linear perturbations on top of a static, spherically symmetric background \citep{ReggWhee57}.  The result was a reduction in the Einstein Field Equations (EFE) that describe spacetime curvature in GR to only five complex-valued equations for the perturbation amplitudes.  Further theoretical work showed this system was only 4th-complex-order, with two degrees of freedom describing the fluid perturbations and two describing the gravitational perturbations \citep{IpseThor73}.  Later, 
\citet[][]{LindDetw85} \citep[see also][]{1983ApJS...53...73L}  
reduced the perturbed EFE to the explicit form of four 1st-order complex-valued equations describing the normal mode perturbations.  For quadrupole modes or higher ($\ell\geq2$), the two gravitational degrees of freedom at the surface produce outgoing gravitational radiation (i.e.~gravitational waves) which will gradually damp any excitations, so that stellar perturbations in GR can be at best quasinormal.  

In asteroseismology, the outgoing gravitational radiation is largely an undesired complication, requiring specialized methods to avoid carrying the boundary condition out to spatial infinity \citep{ChanDetw75, Fack71, LindMendIpse97, AndeKokkSchu95}.  The outgoing gravitational waves can be easily removed using a form of the Cowling approximation within GR, first developed by \citet{1983ApJ...268..837M} and further studied by \citet{LindSpli90}, \citet{1985ApJ...297L..37M}, and \citet{2002A&A...395..201Y}.  In this {\it relativistic Cowling} approximation, the gravitational degrees of freedom are set to zero, retaining only the fluid perturbations.  Further, there is no intrinsic damping, so that the problem becomes real-valued and the modes are stationary.  This treatment is widely used to study the pulsation and stability of compact stellar objects in situations where knowledge of the outgoing gravitational waves is irrelevant, and especially in stars with surface crystallization \citep{2017PhRvC..96f5803F, 2002A&A...395..201Y}.  Another approach to include the relativistic effects in stellar pulsations is to use the {\it Post-Newtonian approximation} \citep{Cutl91, 2014grav.book.....P, 
 2023arXiv230413055R}.  This approach is able to include gravitational perturbations in the form of two scalar potentials and a vector potential, without also producing gravitational radiation \citep{reese..bostonphd}.  

Most interest in pulsations of relativistic stars has focused on neutron stars 
\citep[e.g.][]{McDeVanHHans88, CutlLind92, LindSpli89}.
The earliest calculations of pulsations in WDs involving GR tried to address the origin of radio sources discovered by \citet{1968Natur.217..709H}, as an alternative to a neutron star origin \citep{1968ApJ...152L..71T}. These studies, which date back to the late 1960s, were devoted to computing the fundamental radial pulsation mode of 
Hamada-Salpeter WD models \citep{1961ApJ...134..683H} including GR effects \citep{1968Natur.218..734F, 1968Natur.218..923S, 1969Ap&SS...5..113C}. 
\cite{2023arXiv230413055R} have recently renewed interest in this topic by focusing on relativistic pulsations of ZZ Ceti stars and other pulsating WDs, concentrating on average-mass WDs. In the present paper, we study the impact of GR on realistic evolutionary stellar models of ultra-massive DA WDs computed by  \cite{2022A&A...668A..58A}, which are representative of very high-mass ONe-core ZZ Ceti stars. As a first step, in this work we adopt the relativistic Cowling approximation described above to incorporate relativistic effects in the pulsation calculations, following the treatment provided in \cite{2002A&A...395..201Y}.
In future papers we plan to examine the Post-Newtonian and full 4th-order GR equations, applied to ultra-massive ONe-core WDs and to ultra-massive CO-core WDs (C\'orsico et al.~2023b, in prep.).

The paper is organised as follows. In Sect. \ref{models} we briefly 
describe the relativistic WD models computed by \cite{2022A&A...668A..58A},
emphasising the impact of GR  on the stellar structure. We devote Sect. \ref{RNSP} to describe our approach for the relativistic nonradial stellar pulsations, particularly the 
formalism of the relativistic Cowling approximation (Sect. \ref{RCA}, \ref{rfactors},
\ref{LBVF} and \ref{diffeq}).  The pulsation results for our ultra-massive WD models are described in Sect. \ref{pulsation_results}. Finally, in Sect. \ref{conclusions} we 
summarise our findings. We present in Appendix \ref{Appendix1} a derivation of the relativistic version of the "modified Ledoux" treatment of the Brunt-V\"ais\"al\"a frequency, and in Appendix \ref{Appendix2} the results of a validation of the main results of the paper using a toy model based on Chandrasekhar's models.

\section{Relativistic ultra-massive WD models}
\label{models}

To determine whether to employ GR or Newtonian gravity in a system like a star, a qualitative general criterion commonly used is to assess the magnitude of the "relativistic correction factor", $\varepsilon$, defined as $\varepsilon= GM_{\star}/c^2 R_{\star}$, where $G$ is the Newtonian gravitational constant, $c$ is the speed of light, and $M_{\star}$ and  $R_{\star}$ are the stellar mass and radius, respectively \citep{2014grav.book.....P}\footnote{The parameter $\varepsilon$ is nothing more than the surface gravitational redshift in the Newtonian limit, $z$.}. The larger  $\varepsilon$, the worse the approximation of Newtonian gravity. For instance, for a neutron star, $\varepsilon$ is of order $\sim 0.1$, while for a black hole, $\varepsilon \sim 1$. For average mass ($\sim 0.6 M_{\sun}$) WDs, $\varepsilon$ is $\sim 10^{-4}$, and that is why until recently the relativistic effects have been neglected in the calculation of their structures. If we instead consider an ultra-massive WD star with $M\sim 1.3 M_{\sun}$ and $\varepsilon \sim 0.001$, at first glance, it is not clear if the relativistic effects should be included or not. However, \cite{2018GReGr..50...38C} showed that for the most massive WDs, the importance of GR for their structure and evolution cannot be ignored. In fact, numerous 
works based on static WD structures have shown that GR effects are relevant for the determination of the radius of massive WDs \citep{2011PhRvD..84h4007R, 2017RAA....17...61M, 2018GReGr..50...38C, 2021ApJ...921..138N}. In particular, these studies have demonstrated that for fixed values of mass, deviations of up to $50 \%$ in the Newtonian WD radius are expected compared to the GR WD radius. Recently, \cite{2022A&A...668A..58A} have presented the first set of constant rest-mass ONe-core ultra-massive WD evolutionary models with masses greater than $\sim 1.30 M_{\sun}$ (and up to $1.369 M_{\sun}$) that fully take into account the effects of GR.  This study demonstrates that the GR effects must be considered to assess the structural and evolutionary properties of the most massive WDs. 
This analysis has been extended recently by \cite{2023MNRAS.523.4492A} 
to ultra-massive WDs with CO cores that result from the complete evolution of single progenitor stars that avoid C-ignition \citep{2021A&A...646A..30A,2022MNRAS.511.5198C}.

\cite{2022A&A...668A..58A} employed the {\tt LPCODE} stellar evolution code,  appropriately modified to take into account relativistic effects. They considered  
initial chemical profiles as predicted by the progenitor evolutionary history \citep{2007A&A...476..893S,2010A&A...512A..10S,2019A&A...625A..87C}, and computed model sequences of $1.29, 1.31, 1.33, 1.35$, and $1.369 M_{\odot}$ WDs. The standard equations of stellar structure and evolution were generalised to include the effects of GR following \cite{1977ApJ...212..825T}. In particular, the modified version of {\tt LPCODE} computes the  dimensionless GR correction factors $\mathscr{H, G, V,}$ and $\mathscr{R}$ which turn to unity in the Newtonian limit. These factors correspond, respectively, to the enthalpy, gravitational acceleration, volume, and redshift correction parameters. For comparison purposes, 
\cite{2022A&A...668A..58A} have also computed the same WD sequences for the Newtonian gravity case. All these sequences included the energy released during the crystallisation process, both due to latent heat and the induced chemical redistribution, as in
\cite{2019A&A...625A..87C}. 

\begin{figure}
\includegraphics[width=1.\columnwidth]{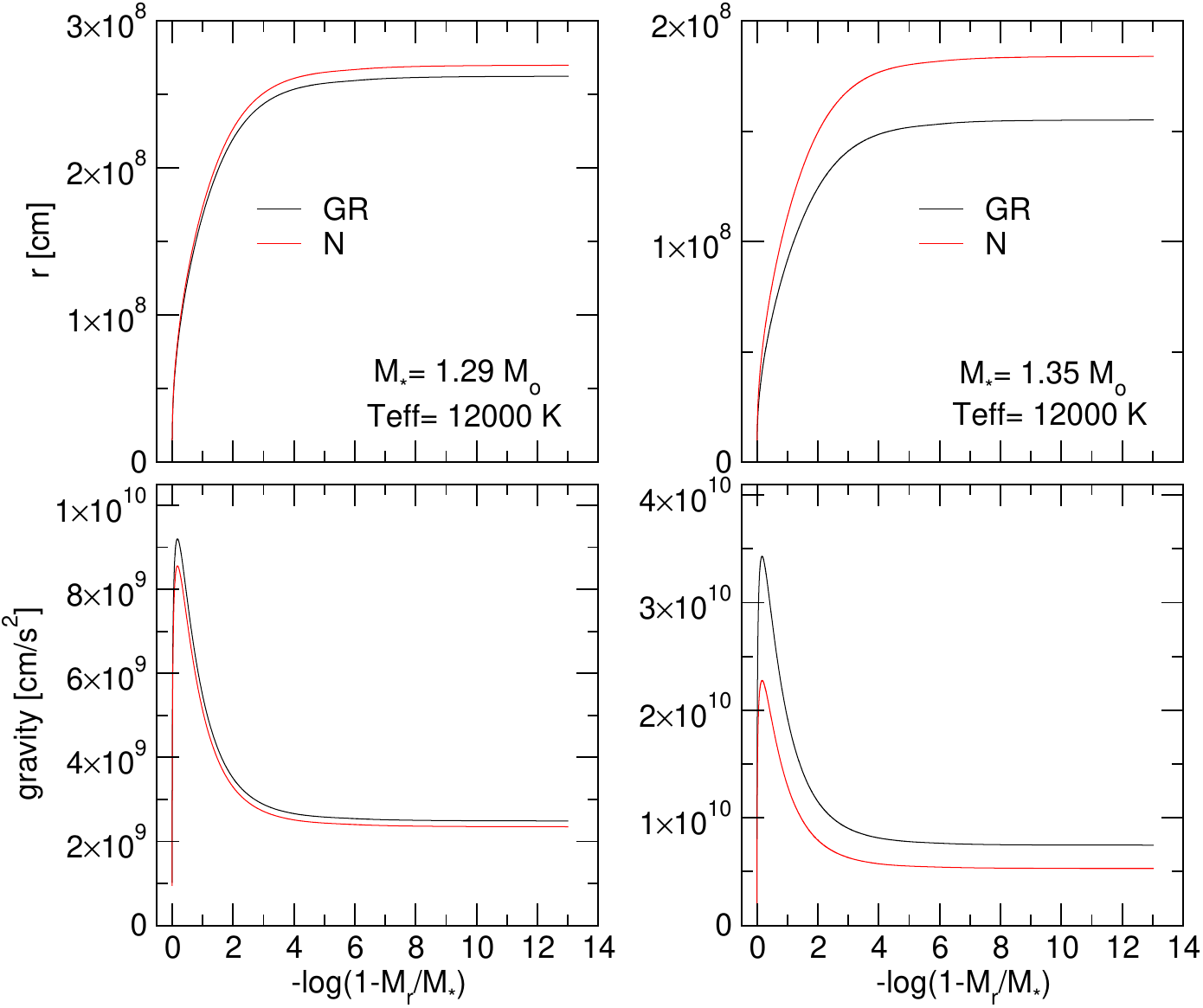}
\caption{The stellar radius (upper panels) and gravity (bottom panels)
    in terms of the outer mass fraction coordinate corresponding to
    ultra-massive DA WD models with $M_{\star}= 1.29 M_{\odot}$ (left) and
    $M_{\star}= 1.35 M_{\odot}$ (right), the GR case (black curves) and for the N case (red curves) ($T_{\rm eff} \sim  12\,000$ K).} 
        \label{fig:radio-gravedad}
\end{figure}

We briefly describe below some of the properties of the representative models of ultra-massive ONe-core WD stars, emphasising the impact of GR on their structure. We refer the reader to the paper by \cite{2022A&A...668A..58A} for a detailed description of the effects of GR on the structural properties of these models. Here, we choose two template WD models characterised by  stellar masses $M_{\star}= 1.29 M_{\sun}$ and $M_{\star}= 1.35 M_{\sun}$, 
H envelope thickness of $\log(M_{\rm H}/M_{\star}) \sim -6$, and an effective temperature of $T_{\rm eff}\sim 12\,000$ K, typical of the ZZ Ceti instability strip. We distinguish two cases: one in which we consider Newtonian WD models (N case), and another one in which the WD structure is relativistic (GR case). In Fig. \ref{fig:radio-gravedad}  we plot the run of the stellar radius and gravity in terms of the outer mass fraction coordinate, corresponding to WD models with $1.29 M_{\sun}$ (left panels) and $1.35 M_{\sun}$ (right panels), for the GR case (black curves) and the N case (red curves).  Clearly, GR induces smaller radii and larger gravities, and this effect is much more pronounced for larger stellar masses. 

\begin{figure}
\includegraphics[width=1.\columnwidth]{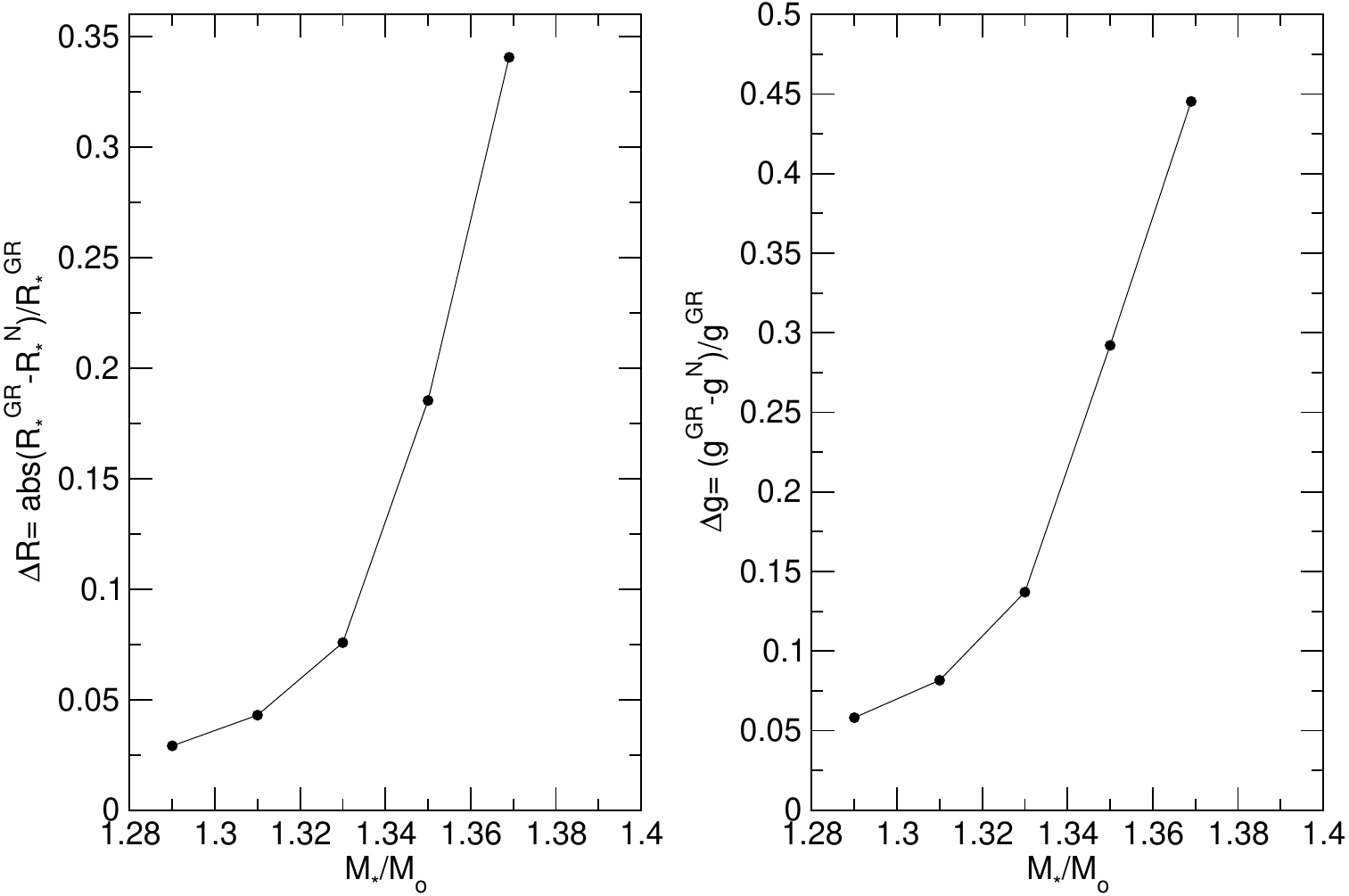}
        \caption{Left: the absolute relative difference between relativistic and Newtonian stellar radius versus stellar mass. Right: the relative difference between the relativistic and Newtonian surface gravity in terms of the stellar mass.} 
        \label{fig:radio_masa}
\end{figure}

In Table \ref{table:M_R}, which is a shortened version of Table 1 of \cite{2022A&A...668A..58A}, we include the values of the stellar radius and the surface gravity for models with $T_{\rm eff}= 10\,000$ K and masses between $1.29 M_{\sun}$ and $1.369 M_{\sun}$ in the GR and N cases. As can be seen, the impact of GR on the radius and gravity of the models is noticeable. In Fig. \ref{fig:radio_masa} we plot the relative differences $\Delta R_{\star}= |R_{\star}^{\rm GR}-R_{\star}^{\rm N}|/R_{\star}^{\rm GR}$ (left panel) and $\Delta g= (g^{\rm GR}-g^{\rm N})/g^{\rm GR}$ (right panel) in terms of the stellar mass. The stellar radius is lower by $\sim 3 \%~(1.29 M_{\sun})$ to $\sim 34\%~ (1.369 M_{\sun})$, and the surface gravity is higher by $\sim 6 \%~ (1.29 M_{\sun})$ to $\sim 44\%~ (1.369 M_{\sun})$ compared to the case where GR is neglected. The typical observational uncertainties in the radii and surface gravities of the most massive WDs in the Montreal White Dwarf Database 100 pc sample \citep{2021MNRAS.503.5397K} are 3\% and 6\%, respectively. Hence, the differences between the GR and N cases can be detected observationally for WDs with masses above $\sim 1.3~M_{\odot}$. These discrepancies must have important consequences for the pulsational properties of ultra-massive WDs, as we will see in Sect. \ref{pulsation-results}.

\begin{table}
\centering
\caption{Stellar masses, radius and surface gravities of the ultra-massive ONe-core WD models at $T_{\rm eff}= 10\,000$ K in the relativistic  and Newtonian cases.}
\begin{tabular}{lcccc}
\hline
\hline
\noalign{\smallskip}
$M_{\star}/M_{\odot}$ & $R_{\star}^{\rm GR}$ &  $R_{\star}^{\rm N}$ & $\log g^{\rm GR}$ & $\log g^{\rm N}$ \\
\noalign{\smallskip}
& [$\times 10^{8}$ cm] &  [$\times 10^{8}$ cm] & [cm/s$^2$] & [cm/s$^2$] \\
\noalign{\smallskip}
\hline
\noalign{\smallskip}
1.29 &  2.609 & 2.685 & 9.401 & 9.375 \\ 
1.31 &  2.326 & 2.426 & 9.507 & 9.470 \\
1.33 &  2.005 & 2.157 & 9.643 & 9.579 \\
1.35 &  1.543 & 1.829 & 9.878 & 9.728 \\
1.369 & 1.051 & 1.409 & 10.217 & 9.961 \\
\hline
\end{tabular}
\label{table:M_R}
\end{table}

\section{Relativistic nonradial stellar pulsations in WDs}
\label{RNSP}

In order to incorporate the relativistic effects in the pulsations of WDs, we adopt the relativistic Cowling approximation in the form developed by \citet{2002A&A...395..201Y}, 
and follow the GR formalism provided in \citet{reese..bostonphd}.

\subsection{The relativistic Cowling approximation}
\label{RCA}

The Cowling approximation of Newtonian nonradial pulsations \citep[named after T. G. Cowling's pioneer paper;][]{1941MNRAS.101..367C} is based on neglecting the gravitational potential perturbations during the fluid oscillations. This approximation has been widely used in Newtonian nonradial pulsation computations in the past, because it constitutes a 2nd-order differential eigenvalue problem, thus simplifying the complete 4th-order problem \citep{1989nos..book.....U}. It is also a very good approximation to periods of $g$-modes in WDs, which are primarily envelope modes \citep{1999ApJ...525..482M}. The Cowling approximation has been frequently used in asymptotic treatment of stellar pulsations \citep[see, for instance,][]{1980ApJS...43..469T}, and also in numerical treatments of $g$-mode pulsations in rapidly rotating WDs \citep[e.g.,][]{2019MNRAS.487.2177S,2023ApJ...951..122K}, although it has fallen out of use in the context of present-day numerical calculations of Newtonian  nonradial stellar pulsations and asteroseismology. The {\it relativistic} Cowling approximation \citep{1983ApJ...268..837M}, on the other hand,  is generally employed in the field of pulsations of relativistic objects such as neutron stars  \citep{1989ApJ...345..925L, 2002A&A...395..201Y,2020PhRvD.102f3025S} and hybrid (hadron plus quark matter phases) neutron stars \citep{2020PhRvD.101l3029T,2023PhRvD.107j3048Z}. 

In the next sections, we first describe the relativistic correction factors involved in the pulsation problem. Then, we provide relativistic expressions to calculate the critical frequencies (Brunt-V\"ais\"al\"a and Lamb frequencies), after which we assess the coefficients of the pulsation differential equations in the relativistic Cowling form. Finally, we provide the two first order differential equations to be solved, along with the boundary conditions of the eigenvalue problem.

\subsection{Relativistic correction factors  $\mathscr{R}$,
  $\mathscr{V}$, and potentials $\nu$, $\lambda$}
\label{rfactors}

We start by considering the Schwarzschild metric of GR for spacetime inside and around a star \citep{1977ApJ...212..825T}:

\begin{equation}
ds^2 = - e^{2\Phi/c^2} c^2 dt^2 + \left(1-\frac{2Gm}{c^2r}\right)^{-1} dr^2 + r^2 d\Omega^2,
\label{eq:dsThorne}  
\end{equation}  

\noindent where $m$ is the "total mass inside radius $r$", which
includes the rest mass, nuclear binding energy, internal energy, and
gravity. $\Phi$ is a gravitational
potential, which in the Newtonian limit $c\rightarrow\infty$ corresponds to the scalar Newtonian potential. 

Following \citet{1977ApJ...212..825T} in his treatment of relativistic stellar interiors, it is convenient to write the metric in the form

\begin{equation}
ds^2 = - \mathscr{R}^2 c^2 dt^2 + \mathscr{V}^2 dr^2 + r^2 d\Omega^2,
\label{eq:dsThorne2}  
\end{equation}  

\noindent where the redshift correction factor $\mathscr{R}$, and
the volume correction factor $\mathscr{V}$, are
defined as \citep{1977ApJ...212..825T}:

\begin{equation}
\mathscr{R}= e^{\Phi/c^2},\ \ \  \mathscr{V}= \left(1-\frac{2Gm}{c^2 r} \right)^{-1/2}.
\label{eq:R_V}  
\end{equation} 

\noindent The metric is usually written also as a function of two relativistic gravitational potentials $\nu$ and $\lambda$  \citep{1939PhRv...55..364T, 1939PhRv...55..374O}, so that:

\begin{equation}
ds^2 = - e^\nu c^2 dt^2 + e^\lambda dr^2 + r^2 d\Omega^2.
\label{eq:dsThorne3}  
\end{equation} 

\noindent Equating Eqs. (\ref{eq:dsThorne}) and (\ref{eq:dsThorne3}), we have

\begin{equation}
\nu= \frac{2\Phi}{c^2},\ \ \ {\rm and}\ \ \ \lambda= -\ln\left(1-\frac{2Gm}{c^2 r} \right).
\label{eq:nu_lamda}  
\end{equation} 

\noindent We obtain $\nu$ and $\lambda$ in terms of the
variables $\mathscr{R}$ and $\mathscr{V}$, that are the output of the 
relativistic {\tt LPCODE} version \citep{2022A&A...668A..58A} by equating Eqs.
(\ref{eq:dsThorne2}) and (\ref{eq:dsThorne3}):

\begin{equation}
\mathscr{R}^2= e^\nu, \ \ \ \mathscr{V}^2= e^\lambda,
\label{eq:R2-V2}  
\end{equation} 

\noindent so that,

\begin{equation}
\nu= 2 \ln \mathscr{R}, \ \ \ \lambda= 2 \ln \mathscr{V}.
\label{eq:nu-lambda}  
\end{equation} 

In the Newtonian limit, we have
$\mathscr{R}= e^{\nu/2} \rightarrow 1$ and
$\mathscr{V}= e^{\lambda/2} \rightarrow 1$,
so that $\nu, \lambda \rightarrow 0$.

We compute the derivatives of $\nu$ and $\lambda$
by calculating the numerical derivatives of $\mathscr{R}$ and $\mathscr{V}$
 as:

\begin{equation}
\nu^{\prime}\equiv\frac{d\nu}{dr}= \frac{2}{\mathscr{R}} \left(\frac{d\mathscr{R}}{dr}\right), \ \ \  \lambda^{\prime} \equiv \frac{d\lambda}{dr}= \frac{2}{\mathscr{V}} \left(\frac{d\mathscr{V}}{dr}\right). 
\label{eq:nu_prima_lambda_prima}  
\end{equation}

\begin{figure}
\includegraphics[width=1.\columnwidth]{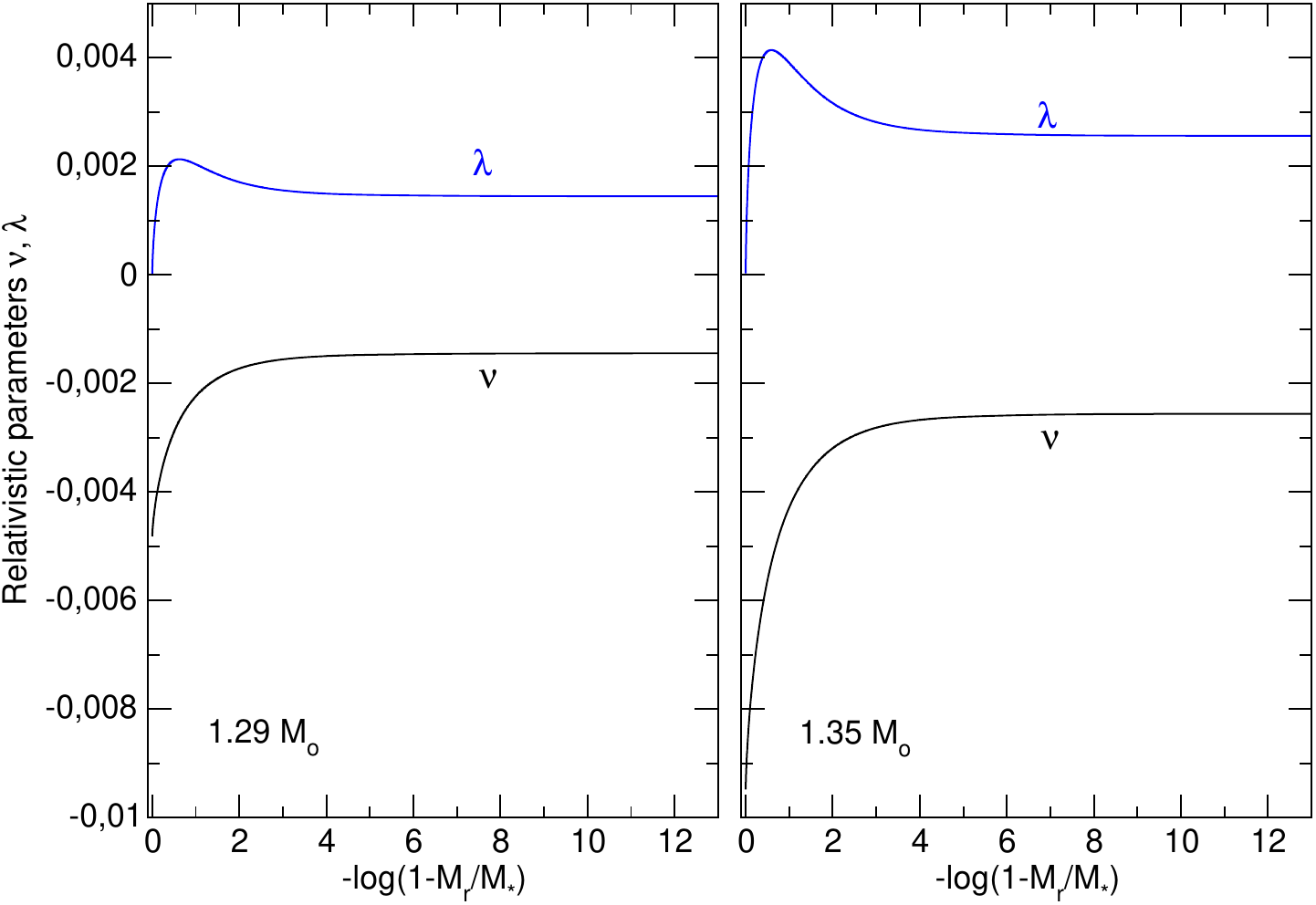}
        \caption{The relativistic potentials $\nu$ (black) and $\lambda$ (blue), in terms 
        of the outer mass fraction coordinate, corresponding to two template ONe-core WD models with masses $M_{\star}= 1.29 M_{\sun}$ and $M_{\star}= 1.35 M_{\sun}$, and effective temperature $T_{\rm eff}\sim 12\,000$ K.  Note the high peaks in the stellar center, indicating the increased gravitational strength of the core, and that the more massive star shows more extreme values.  At the surface, $\nu\rightarrow-\lambda$.}
        \label{fig:nu-lambda}
\end{figure}

\begin{figure}
\includegraphics[width=1.\columnwidth]{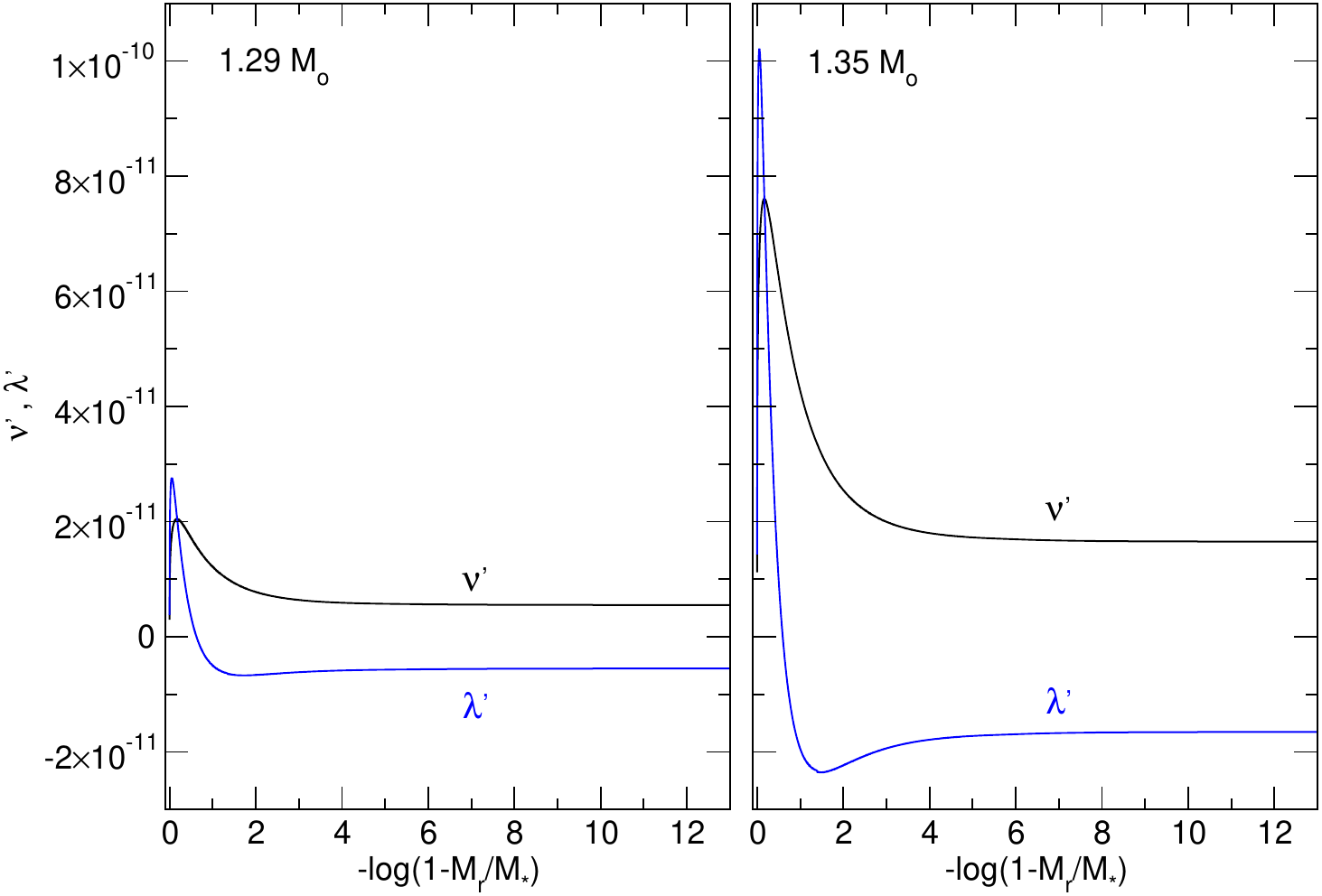}
        \caption{The derivatives of the relativistic potentials $\nu^{\prime}$ (black) and $\lambda^{\prime}$ (blue), in terms of the outer mass fraction coordinate, corresponding to the same template models shown in Fig. \ref{fig:nu-lambda}.  Note again the sharp peaks in the core, and more extreme values in the more masive star.} 
        \label{fig:dnu-dlambda}
\end{figure}

\begin{figure}
\includegraphics[width=1.\columnwidth]{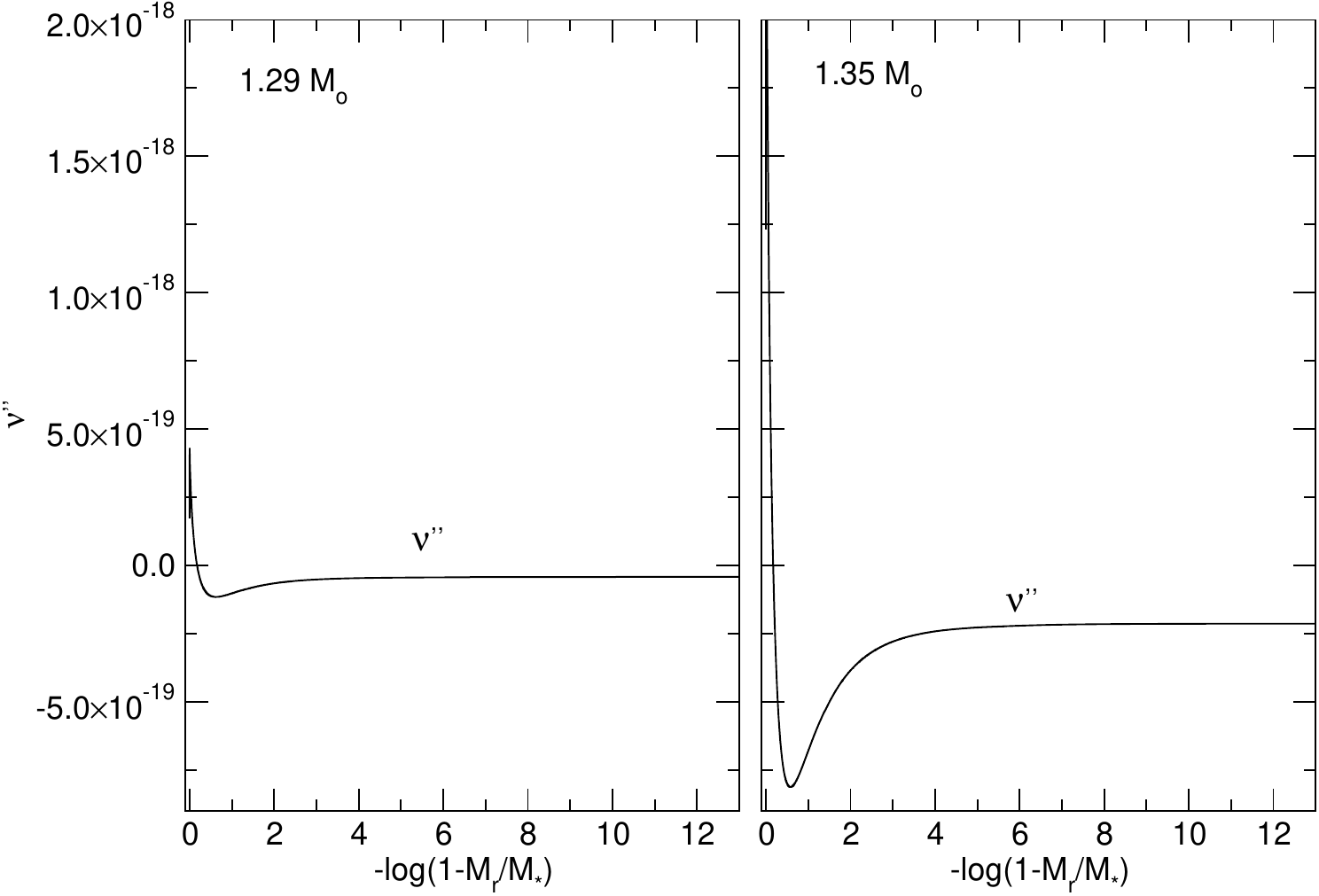}
        \caption{The second derivative of $\nu$, that is, $\nu^{\prime\prime}$, in terms of the outer mass fraction coordinate, corresponding to the same template models shown in Fig. \ref{fig:nu-lambda}.} 
        \label{fig:ddnu}
\end{figure}

\begin{figure*}
\begin{center}
\includegraphics[width=17.0cm]{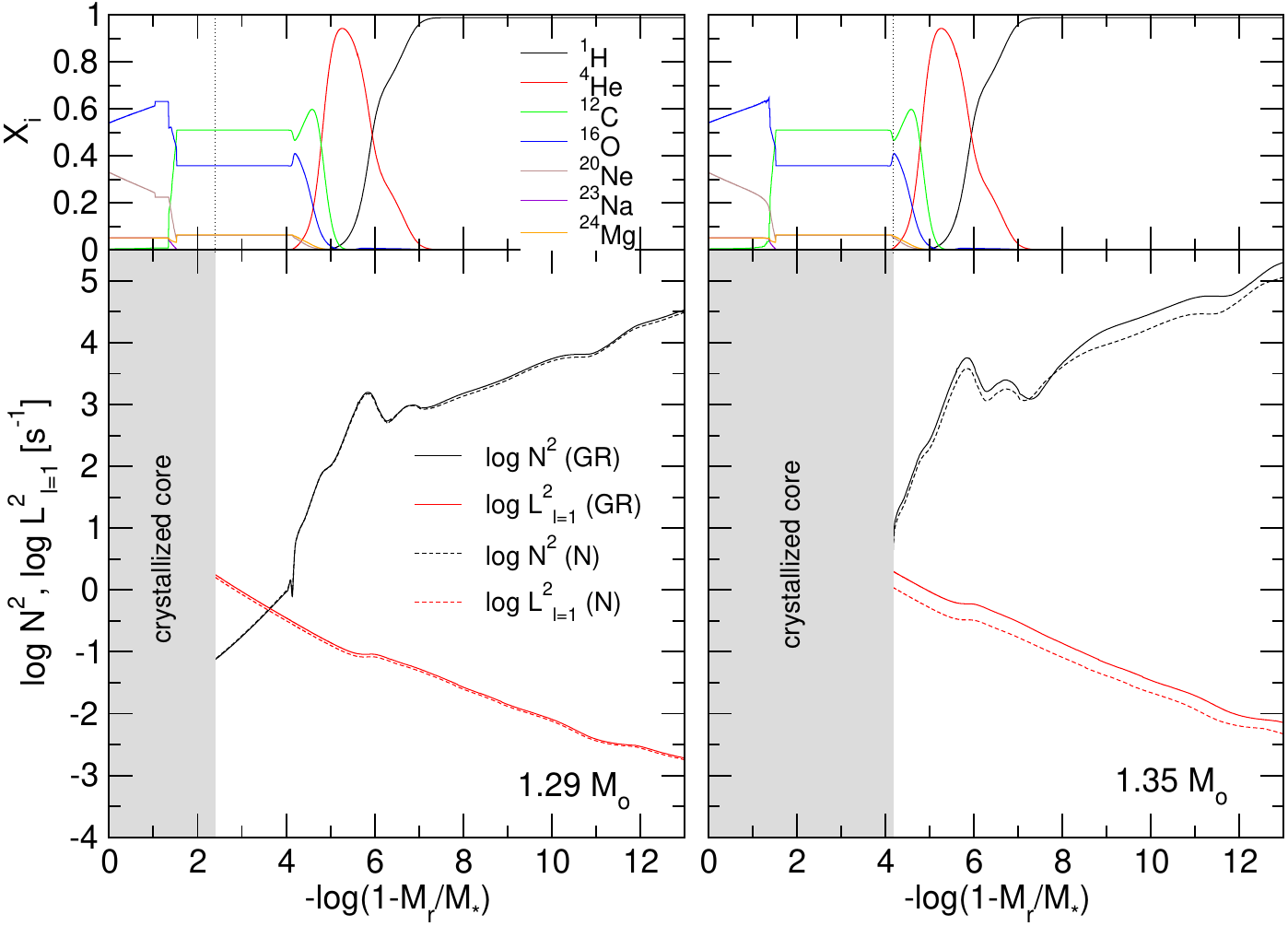}
        \caption{Upper panels: abundances by mass of the different chemical species as a function of the fractional mass, corresponding to the template WD models with masses $M_{\star}= 1.29 M_{\sun}$ (left) and $M_{\star}= 1.35 M_{\sun}$ (right), and effective temperature $T_{\rm eff}\sim 12\,000$ K. Lower panels: logarithm of the squared Brunt-V\"ais\"al\"a and Lamb ($\ell= 1$) frequencies for the GR case (solid lines) and the N case (dashed lines). The grey areas correspond to the crystallised core regions, in which $g$ modes cannot propagate.} 
        \label{fig:Xi_N2}
\end{center}
\end{figure*}

The numerical derivatives of $\mathscr{R}$ and $\mathscr{V}$, as well as $\nu^{\prime}$ and $\lambda^{\prime}$, are usually noisy when computed following Eqs. (\ref{eq:nu_prima_lambda_prima}). To avoid this, we compute
$\nu^{\prime}$  and $\lambda^{\prime}$ by employing solutions to the Einstein field equation for the static, spherically symmetric distribution of matter, given by 
\citet{1939PhRv...55..364T} and \citet{1939PhRv...55..374O} \citep[see also][]{Toop64}:

\begin{equation}
e^{-\lambda} \left(\frac{1}{r^2}-\frac{\lambda^{\prime}}{r}\right)-\frac{1}{r^2}= -\frac{8\pi G}{c^4} \rho c^2, 
\label{b1}  
\end{equation}  
  
\begin{equation}
e^{-\lambda} \left(\frac{1}{r^2}+\frac{\nu^{\prime}}{r}\right)-\frac{1}{r^2}= \frac{8\pi G}{c^4} P, 
\label{b2}  
\end{equation}

\begin{equation}
\frac{1}{2} e^{-\lambda} \left[\nu^{\prime\prime} + \frac{1}{2}\left(\nu^{\prime}-\lambda^{\prime}\right) \left(\nu^{\prime}+\frac{2}{r}  \right)\right]= \frac{8\pi G}{c^4} P,    
\label{b3}
\end{equation}  

\noindent where $P$ is the pressure and $\rho$ is the mass-energy density 
(not just the mass density). With some rearranging, we can write:

\begin{equation}
\lambda^{\prime}= \frac{1}{r}+\left(\frac{8\pi G}{c^2}\rho r-\frac{1}{r}\right)e^{\lambda}
\label{b4}  
\end{equation}  

\begin{equation}
\nu^{\prime}= -\frac{1}{r}+\left(\frac{8\pi G}{c^4} P r+\frac{1}{r}\right)e^{\lambda}
\label{b5}  
\end{equation}  

\begin{equation}
\nu^{\prime\prime}= \frac{16\pi G}{c^4} P e^{\lambda}-\frac{1}{2}\left(\nu^{\prime}-\lambda^{\prime}\right)\left(\nu^{\prime}+\frac{2}{r}\right)\equiv \frac{d^2\nu}{dr^2}
\label{b6}  
\end{equation}  

To summarise, in our numerical treatment we employ Eqs. (\ref{b4}) and
(\ref{b5}) to compute $\lambda^{\prime}$ and  $\nu^{\prime}$ using the value
of $\lambda$ calculated with Eq. (\ref{eq:nu-lambda}). We 
employ  Eq. (\ref{b6}) to assess $\nu^{\prime\prime}$ using $\lambda$, $\lambda^{\prime}$, and $\nu^{\prime}$ derived above.
The quantity $\nu^{\prime\prime}$ is required to compute one of the coefficients of the pulsation differential equations
(Sect. \ref{diffeq}).

\subsection{Relativistic adiabatic exponent, sound speed,
Lamb and Brunt-V\"ais\"al\"a frequencies}
\label{LBVF}

The relativistic adiabatic exponent, defined as $\Gamma_1= \left( \frac{\partial\log P}{\partial\log n}\right)_{\rm ad}$, where $n$ is the baryon number density, can be
expressed as  \citep{1967hea3.conf..259T, 1966ApJ...145..514M}:

\begin{equation}
\Gamma_1= \frac{\rho + (P/c^2)}{P} \left(\frac{\partial P}
      {\partial \rho} \right)_{\rm ad}= \frac{\rho+(P/c^2)}{\rho}
      \left(\frac{\partial \log P}
{\partial \log \rho}\right)_{\rm ad},
\label{eq:gamma1}  
\end{equation} 

\noindent This should be compared with the
Newtonian case, where $\Gamma_1= \left(\frac{\partial\log P}{\partial\log \rho}\right)_{\rm ad}$. The relativistic sound speed, $v_s$,
is given by \citep{Curt50}:

\begin{equation}
v_s^2= \frac{\Gamma_1 P}{\rho+(P/c^2)},
\label{eq:vs}  
\end{equation} 

\noindent whereas in the Newtonian case,
$v_s^2= \left(\frac{\partial P}{\partial \rho}\right)_{\rm ad}= \Gamma_1 \frac{P}{\rho}$. 

The squared Lamb and Brunt-V\"ais\"al\"a critical frequencies of the nonradial stellar pulsations, $L_{\ell}^2$ and $N^2$, can be written as \citep{reese..bostonphd}: 

\begin{equation}
  L_{\ell}^2= \ell(\ell+1) \frac{v_s^2}{r^2},
\label{eq:lamb}  
\end{equation} 

\begin{equation}
  N^2= \frac{c^2}{2r} \nu^{\prime} e^{-\lambda}
  \left[\frac{1}{\Gamma_1} \left(\frac{d\log P}{d\log r}\right) -
    \frac{\rho}{\rho+(P/c^2)} \left(\frac{d\log \rho}{d\log r}\right) \right].
\label{eq:bv_def}  
\end{equation} 

\noindent This expression for $N^2$ is analogous to the Newtonian
version of $N^2$ for $g \rightarrow \frac{c^2}{2} \nu^{\prime}$
and an additional relativistic correction factor $e^{-\lambda}$.

The relativistic prescription given by Eq. (\ref{eq:bv_def}) for the assessment of the
Brunt-V\"ais\"al\"a frequency is not well-defined numerically, due to
the high degree of electronic degeneracy prevailing in the core of the
ultra-massive WDs, similar to the case of Newtonian pulsations \citep{1991ApJ...367..601B}. In particular,
the use of Eq. (\ref{eq:bv_def}) leads to unacceptable numerical noise of $N$,  which can lead to miscalculations of the adiabatic $g$-mode periods. To avoid this problem, we employ a numerically convenient relativistic expression, analogous to the Newtonian recipe known as the {\it modified Ledoux} prescription  \citep{1990ApJS...72..335T}. The appropriate relativistic expression for $N^2$, which is derived in Appendix \ref{Appendix1}, is:

\begin{equation}
N^2= e^{-\lambda}\left(\frac{c^2}{2} \nu^{\prime}\right)^2  \frac{\rho+(P/c^2)}{P} 
\frac{\chi_{\rm T}}{\chi_{n}}  \left[\nabla_{\rm ad}-\nabla+B \right],
\label{BVFRel}
\end{equation}

\noindent where $B$ is the Ledoux term, defined as:

\begin{equation}
B= -\frac{1}{\chi_{\rm T}} \sum_{i= 1}^{M-1} \chi_{X_i} \frac{d\ln X_i}{d\ln P},
\label{B_Ledoux}
\end{equation}

\noindent $M$ being the number of different atomic species with fractional abundances $X_i$ that satisfy the constraint $\sum_{i=1}^{M-1} X_i + X_{M}= 1$. The compressibilities $\chi_{\rm T}$, $\chi_{n}$, and $\chi_{X_i}$ are defined as, similar to the Newtonian problem:

\begin{equation}
\chi_{\rm T}= \left(\frac{\partial \ln P}{\partial \ln T}\right)_{n, \{X_i\}},
\ \ \ \chi_{n}= \left(\frac{\partial \ln P}{\partial \ln n}\right)_{T, \{X_i\}},
\ \ \ \chi_{\rm X_i}= \left(\frac{\partial \ln P}{\partial \ln X_i}\right)_{T, n}.
\label{chi_T_chi_n_chi_Xi}
\end{equation}

\noindent Using $(d\ln \rho/d\ln n)= (\rho+P/c^2)/\rho$ 
\citep[Eq. 5.90 of][see also \citealt{1967hea3.conf..259T}]{reese..bostonphd}, the compressibility $\chi_{n}$ can be computed as:

\begin{equation}
\chi_n= \frac{\rho+(P/c^2)}{\rho} \chi_{\rho},
\label{chi_n_rho}
\end{equation}

\noindent where $\chi_{\rho}= \left(\frac{\partial \ln P}{\partial \ln \rho}\right)_{T, \{X_i\}}$. Here, $\nabla_{\rm ad}$ and $\nabla$ are the adiabatic and actual temperature gradients, respectively, defined as:

\begin{equation}
\nabla_{\rm ad}= \left(\frac{\partial \ln T}{\partial \ln P}\right)_{{\rm ad}, \{X_i\}}, \ \ \ \nabla= \frac{d \ln T}{d \ln P}~.
\label{eq:nablas}
\end{equation}

Eq. (\ref{BVFRel}) is completely analogous to the Newtonian expression for the squared Brunt-V\"ais\"al\"a frequency, 
$N^2= g^2 (\rho/P)(\chi_{\rm T}/\chi_{\rho})\left[\nabla_{\rm ad}-\nabla+B\right]$. In the relativistic formula, $g$ has been replaced by $c^2 \nu^{\prime}/2$, and the 
ratio $\rho/P$ becomes $(\rho+P/c^2)/P$. 
There is an additional relativistic factor, $e^{-\lambda}$, 
and the compressibility $\chi_{\rho}$ is replaced 
by $\chi_n$, where $n$ is the baryonic number density. 

\subsection{Differential equations of the relativistic
  Cowling approximation}
\label{diffeq}

Here,  we formulate the system of differential equations of the nonradial pulsations in the relativistic Cowling approximation form, that results when we ignore Eulerian metric perturbations in the pulsation equations \citep{1983ApJ...268..837M}.  This reduces the 4th-complex-order problem of nonradial pulsations in GR to a 2nd-real-order problem, which can be written as two real, 1st-order differential equations. Following \cite{2002A&A...395..201Y}, we define the dimensionless variables $\omega$,  $y_1$, and $y_2$, analogous to Dziembowski's variables in Newtonian pulsations \citep{1971AcA....21..289D}%
\footnote{At variance with \cite{reese..bostonphd}, we use $\sigma$ for the physically meaningful oscillation frequency,  and $\omega$  for the dimensionless frequency, following \cite{1989nos..book.....U}.}:

\begin{equation} 
\omega^2= \frac{R_{\star}^3}{GM_{\star}} \sigma^2, \ \ \
y_1= \frac{\xi_r}{r}e^{-i\sigma t}, \ \ \
y_2= \xi_h C_1 \omega^2 e^{-i\sigma t},
\label{eq:omega_y1_y2}  
\end{equation} 

\noindent where $\xi_r$ and $\xi_h$ correspond to the Lagrangian radial and horizontal displacements, respectively. We also define the following dimensionless functions, analogous to Dziembowski's coefficients \citep{1971AcA....21..289D}, calculated with respect to the
stellar equilibrium model \citep{reese..bostonphd}:

\begin{equation}
  V_g (r)= -\frac{1}{\Gamma_1} \left(\frac{d\log P}{d\log r}\right)=
  \frac{\rho+(P/c^2)}{\Gamma_1 P} \frac{rc^2}{2} \nu^{\prime}, 
\label{eq:vg}  
\end{equation} 

\begin{equation}
U_1 (r)= 2 + r \frac{\nu^{\prime\prime}}{\nu^{\prime}},
\label{eq:u1}  
\end{equation} 

\begin{equation}
U_2 (r)= r \lambda^{\prime},
\label{eq:u2}  
\end{equation} 

\begin{equation}
C_1 (r)= \frac{GM_{\star}}{c^2R_{\star}^3} \frac{2r}{\nu^{\prime}} e^{-\nu},
\label{eq:c1}  
\end{equation} 

\noindent and \citep{Thor66}

\begin{equation}
  A^{\ast}(r) = \frac{1}{\Gamma_1} \left(\frac{d\log P}{d\log r}\right) -
  \frac{\rho}{\rho+(P/c^2)}
  \left(\frac{d\log\rho}{d\log r}\right)= \frac{r N^2}{c^2\nu'/2} e^{\lambda}.
\label{eq:a*}  
\end{equation} 

\noindent In the Newtonian limit, $A^{\ast}$, $V_g$, and $C_1$
will limit to their conventional expressions \citep{1989nos..book.....U}.  On the other hand, in the Newtonian limit we have that $U_1$ tends to $U$,
which is defined in \citet{1989nos..book.....U}, and $U_2 \rightarrow 0$. Using these definitions, and defining $x=r/R_{\star}$, the resulting differential equations for the relativistic Cowling approximation
\citep{1983ApJ...268..837M, 1989ApJ...345..925L,
  2002A&A...395..201Y,reese..bostonphd} are:

\begin{equation}
  x\frac{dy_1}{dx}= \left(V_g-3+U_2 \right) y_1 +
  \left(\frac{\ell(\ell+1)}{C_1 \omega^2} -V_g\right) y_2,
\label{eq:cow1}  
\end{equation} 

\begin{equation}
  x\frac{dy_2}{dx}= \left(e^\lambda C_1 \omega^2 -A^{\ast} \right) y_1 +
  \left(1+A^{\ast}-U_1 \right) y_2.
  \label{eq:cow2}  
\end{equation} 

\noindent In the Newtonian limit, we have $e^\lambda \rightarrow 1$
and $U_2 \rightarrow  0$, and the equations adopt exactly the
form of the Newtonian Cowling approximation \citep{1941MNRAS.101..367C,
1989nos..book.....U}. The boundary conditions for this system of differential 
equations are, at the stellar (fluid) center $(x= 0)$:

\begin{equation}
y_1 C_1 \omega^2 - \ell y_2= 0,
\label{eq:cb1}  
\end{equation} 

\noindent and at the stellar surface ($x= 1$):

\begin{equation}
y_1 - y_2= 0, \ \ \ {\rm and}\ \ \ y_1= 1 \ \ ({\rm normalisation\ condition}).
\label{eq:cb2-cb3}  
\end{equation} 

\noindent These are the same boundary conditions as for the Newtonian Cowling approximation. 

For the ulta-massive WD models considered in this work, the stellar core is crystallised, so that the so called ``hard-sphere'' boundary conditions  \citep{1999ApJ...525..482M} 
may be adopted, which exclude the $g$-mode oscillations from the solid core regions. In that case, Eq. (\ref{eq:cb1}) is replaced by the condition: 

\begin{equation}
y_1= 0\ \ \ {\rm and}\ \ \  y_2=\ {\rm arbitrary} 
\label{eq:cb-hsphere} 
\end{equation}

\noindent at the radial shell $x= x_{\rm crys}$ associated with the outward-moving crystallisation  front, instead of the center of the star ($x= 0$). 
To maintain consistency between Newtonian and GR calculations and for a clean comparison, we assume the same internal boundary condition for the GR case as for the N case, that the eigenfunctions are approximately zero in the solid core, and can be treated with a hard-sphere boundary condition.

In this work, to take into account the relativistic effects on $g$-mode pulsations of crystallised ultra-massive WD models, the {\tt LP-PUL} pulsation code \citep{2006A&A...454..863C} has been appropriately modified to solve the problem of relativistic pulsations in the Cowling approximation as given by Eqs. (\ref{eq:cow1}) and (\ref{eq:cow2}), with boundary conditions given by Eqs. (\ref{eq:cb2-cb3}) and (\ref{eq:cb-hsphere}).

\begin{figure}
\includegraphics[width=1.\columnwidth]{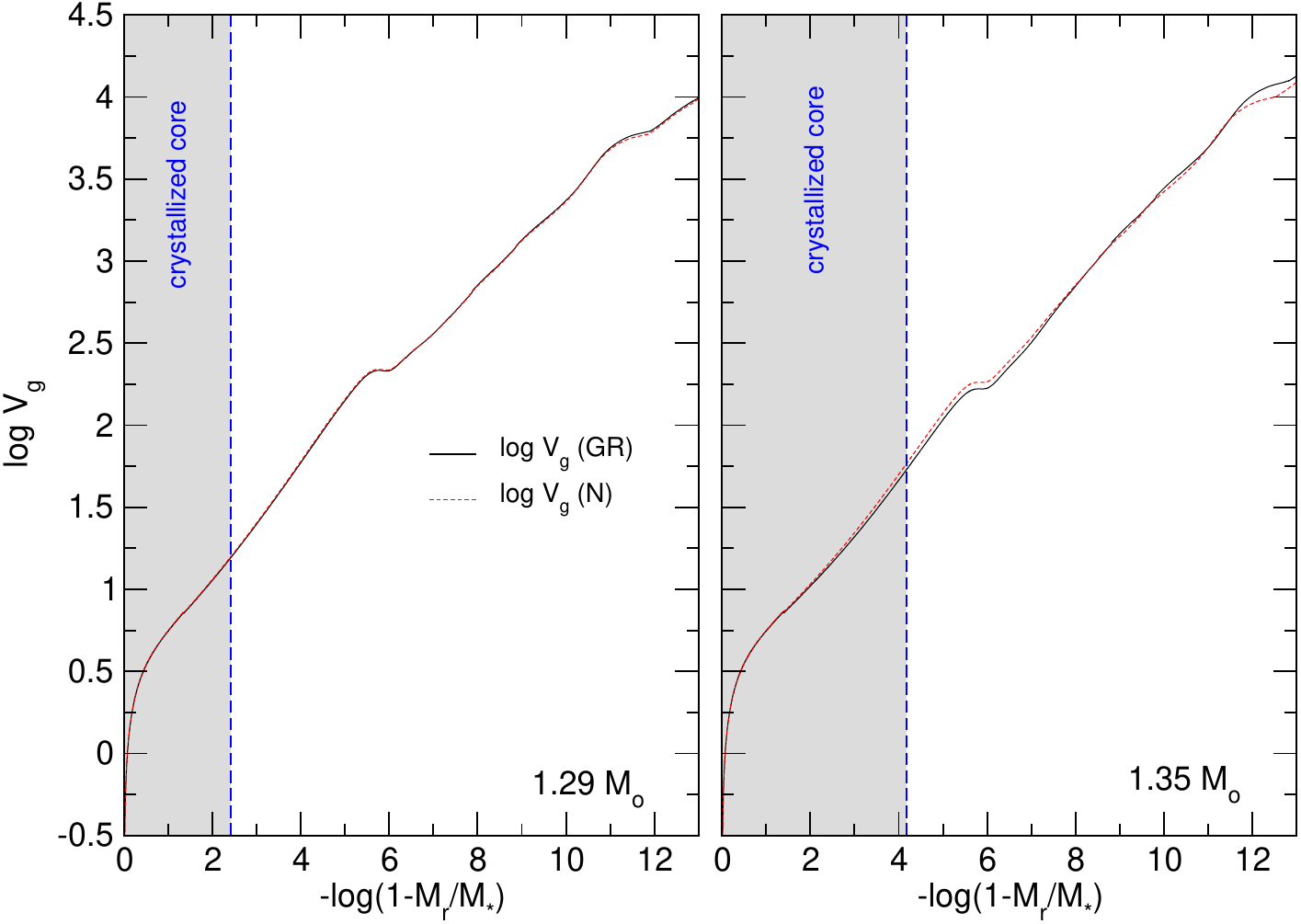}
        \caption{Logarithm of the quantity $V_g$ for the GR case (black solid lines), defined by Eq. (\ref{eq:vg}), as a function of the fractional mass, corresponding to the template WD models with masses $M_{\star}= 1.29 M_{\sun}$ (left) and $M_{\star}= 1.35 M_{\sun}$ (right), and effective temperature $T_{\rm eff}\sim 12\,000$ K. For comparison, we include the function $V_g$ computed for the N case (red dashed lines). The vertical dashed blue line indicates the location of the boundary of the crystallised core region (grey zone).} 
        \label{fig:Vg}
\end{figure}

\begin{figure}
\includegraphics[width=1.\columnwidth]{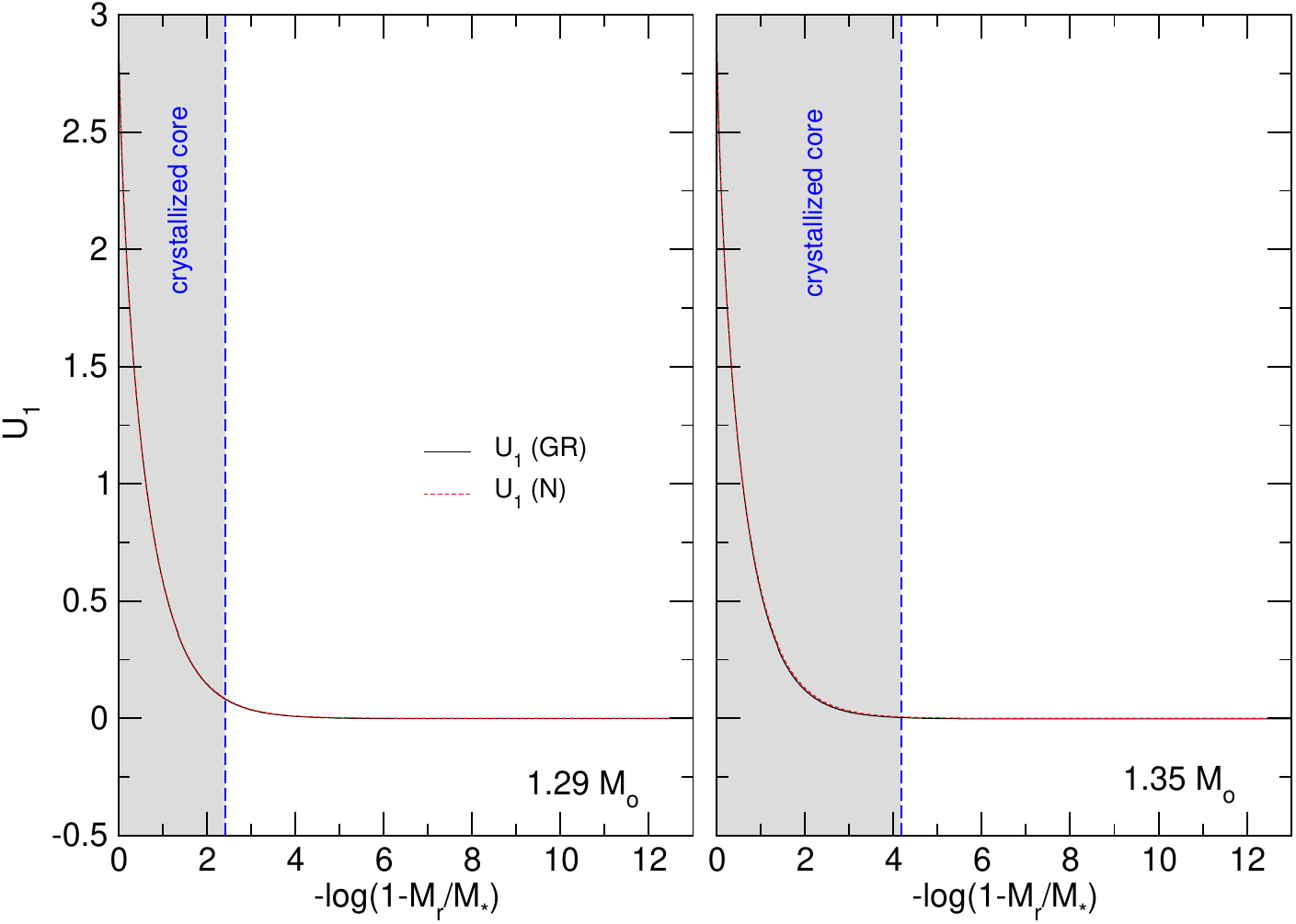}
        \caption{Same as Fig.~\ref{fig:Vg}, but for the quantity $U_1$ (Eq.~\ref{eq:u1}).} 
        \label{fig:U1}
\end{figure}

\begin{figure}
\includegraphics[width=1.\columnwidth]{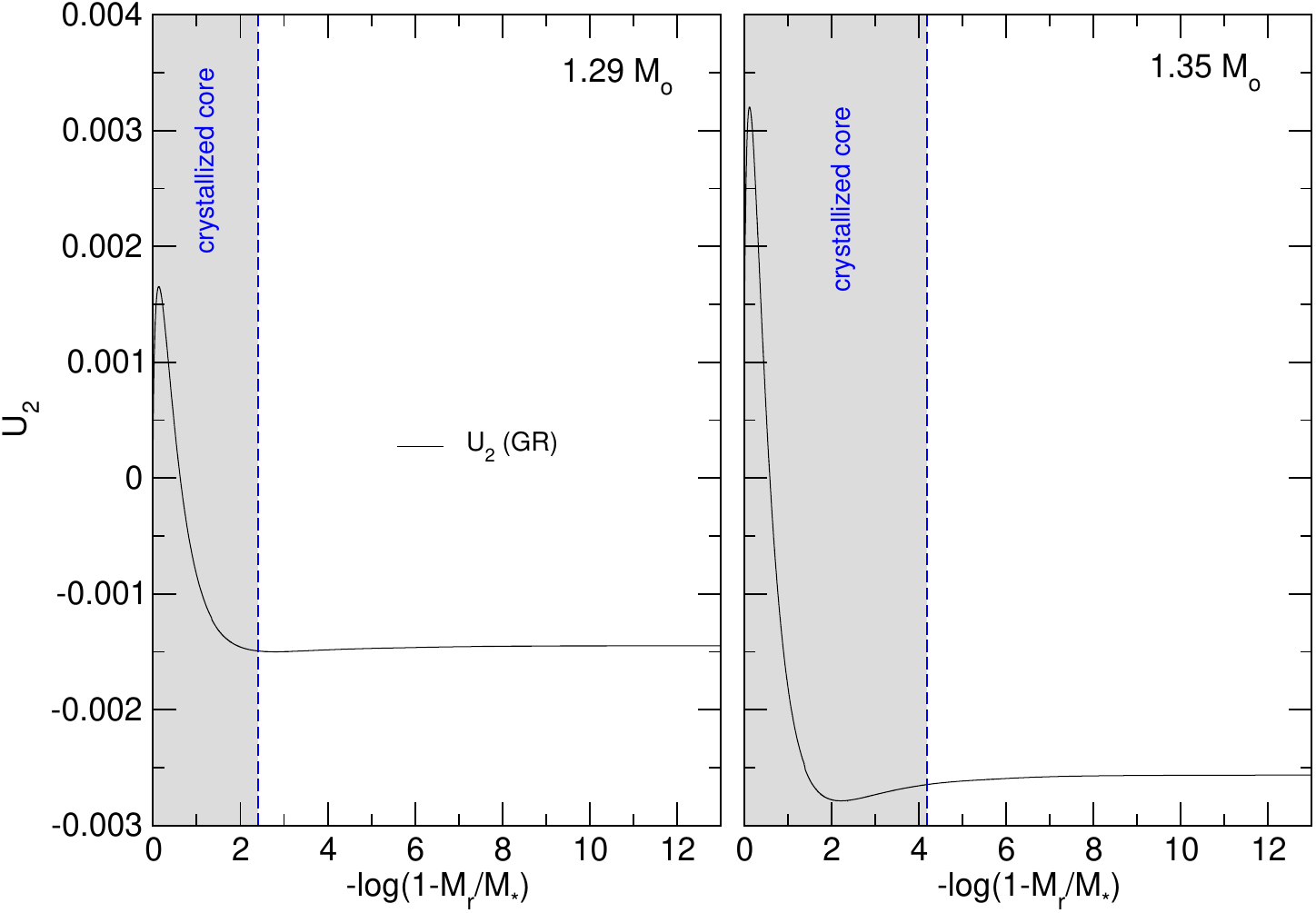}
        \caption{Same as Fig.~\ref{fig:Vg}, but for the quantity $U_2$ (Eq.~\ref{eq:u2}). Since $U_2$ has no counterpart
in the Newtonian pulsation equations, only the GR case  is plotted (black curves).} 
        \label{fig:U2}
\end{figure}

\begin{figure}
\includegraphics[width=1.\columnwidth]{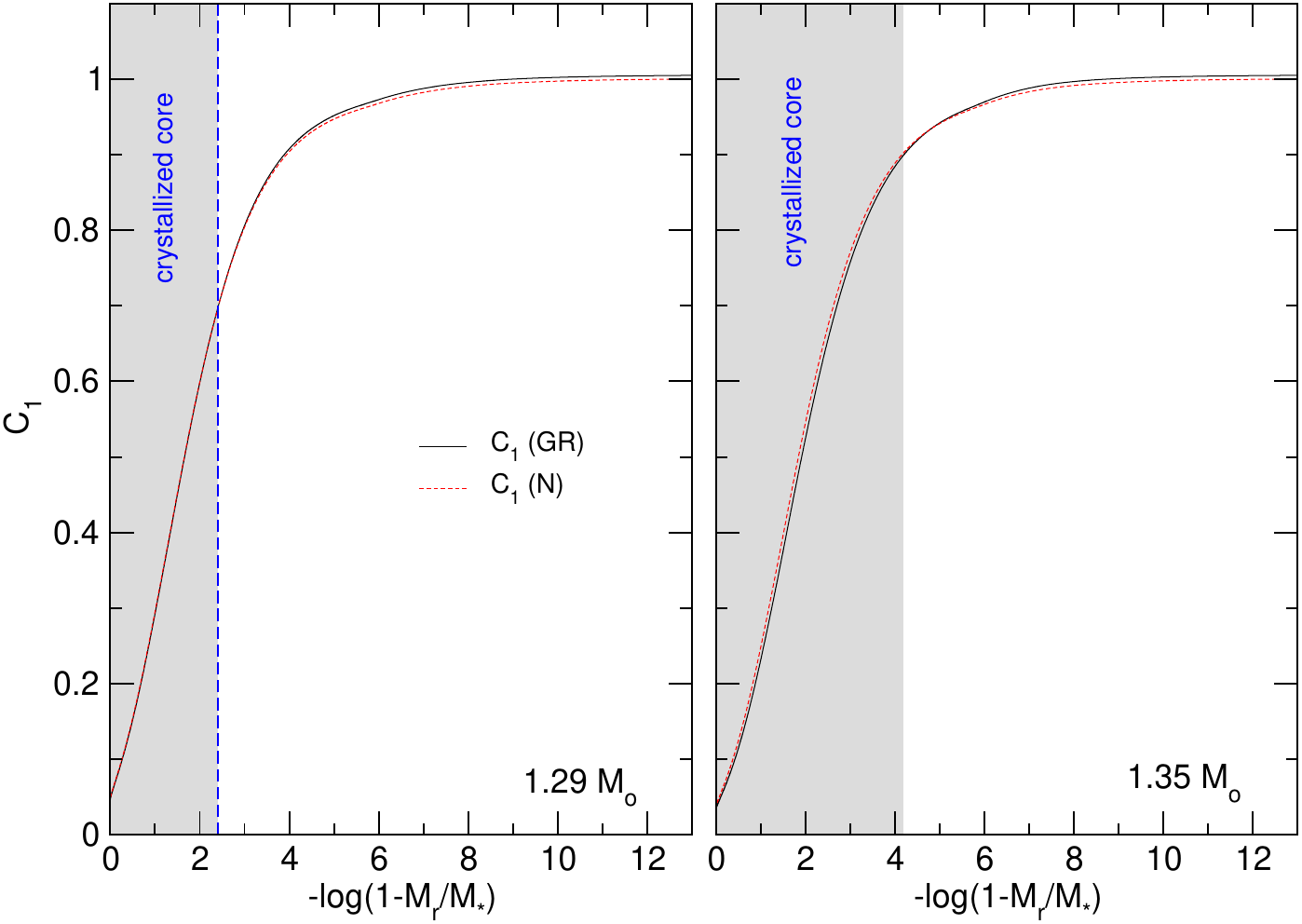}
        \caption{Same as Fig.~\ref{fig:Vg}, but for the quantity $C_1$ (Eq.~\ref{eq:c1}).} 
        \label{fig:C1}
\end{figure}

\begin{figure}
\includegraphics[width=1.\columnwidth]{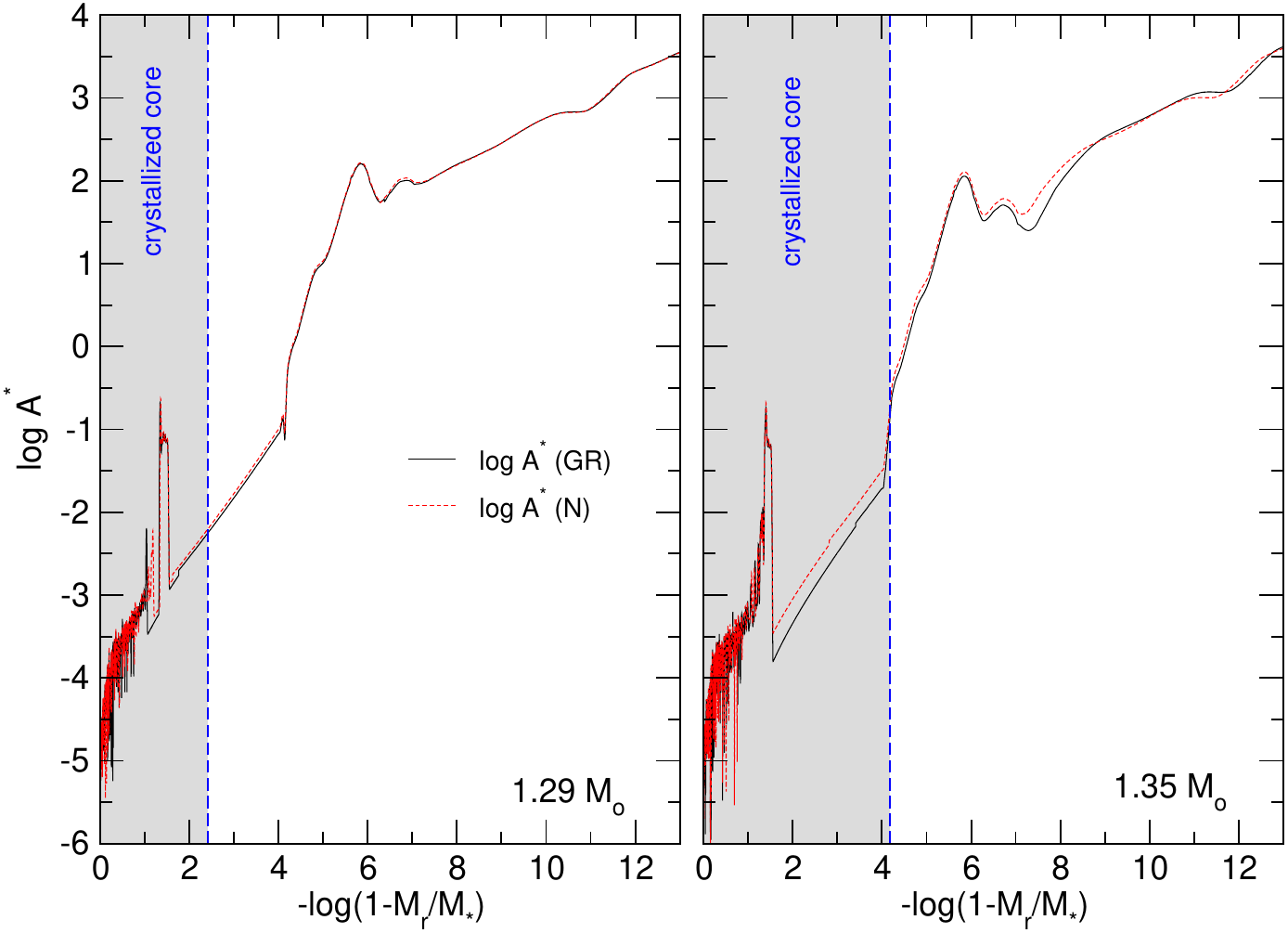}
        \caption{Same as Fig.~\ref{fig:Vg}, but for the logarithm of the quantity $A^*$ (Eq.~\ref{eq:a*}).} 
        \label{fig:A}
\end{figure}

\section{Pulsation results}
\label{pulsation_results}

\subsection{Properties of template models}
\label{models_tools}

It is illustrative to examine the metric parameters $\nu$, $\lambda$, $\nu^{\prime}$, $\lambda^{\prime}$ and $\nu^{\prime\prime}$. In Figs. \ref{fig:nu-lambda}, \ref{fig:dnu-dlambda}, and \ref{fig:ddnu} we show the  $\nu$ and $\lambda$ and their derivatives $\nu^{\prime}, \lambda^{\prime}$ and $\nu^{\prime\prime}$, in terms of the outer mass fraction coordinate, corresponding to the two template WD models with masses $M_{\star}= 1.29 M_{\sun}$ (left) and $M_{\star}= 1.35 M_{\sun}$ (right), and effective temperature $T_{\rm eff}\sim 12\,000$ K. As can be seen, $\nu,\lambda$ quantities are very small throughout the star, being of a similar order with $\varepsilon\sim0.001.$  However, in the center the gravitational values are more extreme than near the surface, pointing to the high central concentration of the mass of these stars.  

The chemical profiles (abundances by mass, $X_i$) of the different nuclear species corresponding to the template models are plotted in the upper panels of Fig. \ref{fig:Xi_N2} as a function of the fractional outer mass. In the lower panels,  we depict the logarithm of the squared Brunt-V\"ais\"al\"a (black lines) and dipole ($\ell= 1$) Lamb (red lines) frequencies for the GR case (solid lines) and the N case (dashed lines). We have emphasised the crystallised regions of the core with grey. The chemical interface of $^{12}$C,$^{16}$O, and $^{20}$Ne, which is located at $-\log(1-M_r/M_{\star}) \sim 1.5$, is embedded in the crystalline part of the core for both template models. Since we assume that $g$-mode eigenfunctions cannot penetrate the solid regions (due to the hard-sphere boundary condition,  
Eq.~\ref{eq:cb-hsphere}), this chemical interface is not relevant for the mode-trapping properties of the models. The chemical transition region between $^{12}$C, $^{16}$O, and $^4$He [$-\log(1-M_r/M_{\star}) \sim 4.5$], which is located in the fluid region in both models, also does not have a significant impact on the mode-trapping properties. Thus, mode-trapping properties are almost entirely determined by the presence of the $^4$He/$^1$H transition, which is located in the fluid external regions, at $-\log(1-M_r/M_{\star}) \sim 6$.  

By closely inspecting Fig. \ref{fig:Xi_N2}, we conclude  that the Brunt-V\"ais\"al\"a and Lamb frequencies for the N and GR cases are similar for the model with $1.29 M_{\sun}$, although they are significantly different for the $M_{\star}= 1.35 M_{\sun}$ model, with both critical frequencies being higher for the GR case than for the N case. Because of this, it is expected that $g$-mode frequencies shift to larger values so that all periods experience a global offset towards shorter values in the relativistic case, compared to the Newtonian case. This will be verified with the calculations of the $g$-mode period spectra in both situations (Sect. \ref{pulsation-results}).

We close this section by comparing the coefficients of the differential equations of the relativistic Cowling approximation with their Newtonian counterparts. In Figs. \ref{fig:Vg} to \ref{fig:A} we depict with black curves the dimensionless functions $V_g, U_1, U_2, C_1$ and $A^*$ in the GR case, as defined by 
Eqs.~(\ref{eq:vg}) to (\ref{eq:a*}), along with the same quantities corresponding to the N case (red curves), computed according to their definition \citep[see, e.g.,][]{1989nos..book.....U}. We include the cases of the two template WD models with $M_{\star}= 1.29 M_{\odot}$ (left panel) and $M_{\star}= 1.35 M_{\odot}$ (right panel). We marked the crystallised region in each model with a grey area with a dashed blue boundary. These figures demonstrate that the dimensionless quantities in the GR case are very similar to the ones for the N case, and this is true for both of the representative models. This is not surprising, since the relativistic correction factors $\nu$ and $\lambda$ and their derivatives $\nu^{\prime}, \lambda^{\prime}$ and $\nu^{\prime\prime}$, that are included in the calculation of the dimensionless coefficients, are small. For the specific case of $A^*$, some numerical noise is observed in the core regions. This is irrelevant for the purposes of this investigation, since those regions are contained in the crystallised core and do not affect the $g$ modes, which are prevented from propagating in the solid phase. 

\subsection{Newtonian and relativistic $g$-mode period spectra}
\label{pulsation-results}

We computed N and GR nonradial $g$-mode  $\ell= 1$ adiabatic pulsation periods in the range $50 \lesssim \Pi \lesssim 2000$ s using an updated version of the {\tt LP-PUL} pulsation code that includes the capability to solve the pulsations equations in the relativistic Cowling approximation described in Sect.~\ref{RCA}. 
The N-case pulsation periods were calculated by solving the differential problem of the Newtonian nonradial  stellar pulsations \citep{1989nos..book.....U}.   We emphasise that in the GR case we are using evolutionary WD models calculated in GR with relativistic, 2nd-order Cowling mode pulsations that ignore gravitational (i.e.~spacetime) perturbations,
while in the N case we are using evolutionary WD models calculated with Newtonian gravity and Newtonian, 4th-order mode pulsations that include gravitational perturbations\footnote{This is at variance with the preliminary results presented in
\cite{ 2023arXiv230204100C}, in which Newtonian equations were used for the $g$-modes, with a fully relativistic WD as the background.}. We have also computed Newtonian periods by solving the 2nd-order Newtonian Cowling approximation \citep{1989nos..book.....U}.  For $g$-modes, these 2nd-order periods are sufficiently similar to the 4th-order periods used in the N case, that the results are not impacted.

In the analysis below, to study the dependence of the relativistic effects on the stellar mass, we compare the $g$-mode period spectra calculated according to the N and GR cases for ultra-massive WD models of different stellar masses at effective temperatures typical of the ZZ Ceti instability strip.

\begin{figure}
\includegraphics[width=1.\columnwidth]{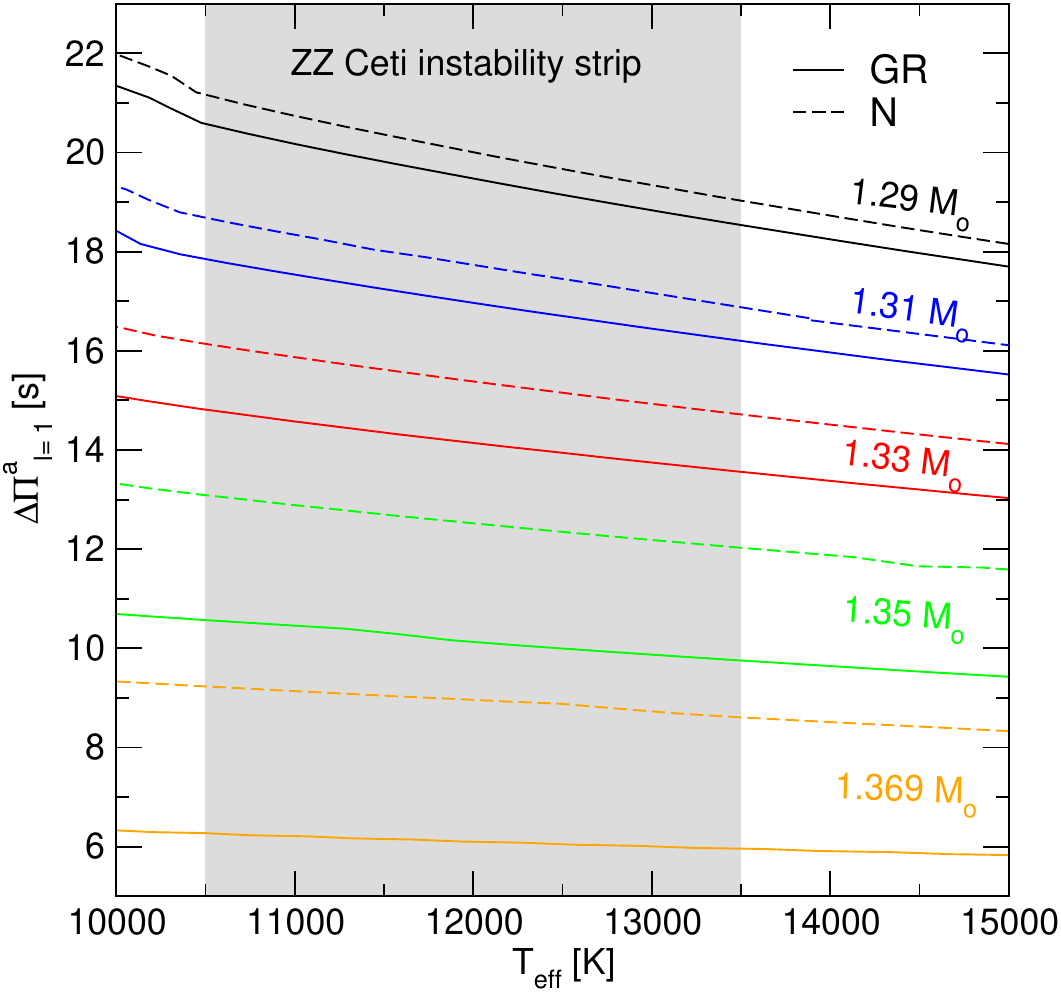}
        \caption{Dipole $\ell= 1$ asymptotic period spacing for ultra-massive WD sequences with different stellar masses for the relativistic (GR) and Newtonian (N) cases in terms of $T_{\rm eff}$ through the whole ZZ Ceti instability strip (light grey area).} 
        \label{fig:asymptotic}
\end{figure}

Before analysing the behaviour of the periods, we first examine the impact of GR on the period spacing of $g$ modes. According to the asymptotic theory of stellar pulsations, and in the absence of chemical gradients, the pulsation periods of the $g$ modes  with  high radial order  $k$  (long  periods) are expected to be uniformly spaced with a
constant period separation given by \citep{1980ApJS...43..469T,1990ApJS...72..335T}:

\begin{equation} 
\Delta \Pi_{\ell}^{\rm a}= \Pi_0 / \sqrt{\ell(\ell+1)},  
\label{eq:aps}
\end{equation}

\noindent where

\begin{equation}
\Pi_0= 2 \pi^2 \left[\int_{\rm fluid} \frac{N}{r} dr\right]^{-1},
\label{eq:pcero}
\end{equation}

\noindent with the integral in Eq. (\ref{eq:pcero}) calculated only in the fluid part of the star. Fig. \ref{fig:asymptotic} depicts
the asymptotic period spacing  for the sequences
of $1.29, 1.31, 1.33, 1.35$ and $1.369~M_{\odot}$ WD models in terms of the effective temperature along the ZZ Ceti instability strip. We find that $\Delta \Pi_{\ell}^{\rm a}$, the asymptotic period spacing, is smaller for the relativistic WD sequences compared to the Newtonian sequences. This is expected, since the asymptotic period spacing is inversely proportional to the integral of the Brunt-V\"ais\"al\"a frequency divided
by the radius. Since the Brunt-V\"ais\"al\"a frequency is larger for the relativistic case (see Fig. \ref{fig:Xi_N2}), the integral is larger and its inverse is smaller than in the Newtonian case. The differences of $\Delta \Pi_{\ell}^{\rm a}$ between the GR and the N cases are larger for higher stellar masses, reaching a minimum difference of $\sim 0.6$~s (which represents a relative variation in period spacing of $\sim 3\%$) for $1.29~M_{\sun}$, and a maximum difference of $\sim 3$ s (that constitutes a relative variation of $\sim 48\%$) for $1.369~M_{\sun}$ for effective temperatures within the ZZ Ceti instability strip.

\begin{figure}
\includegraphics[width=1.\columnwidth]{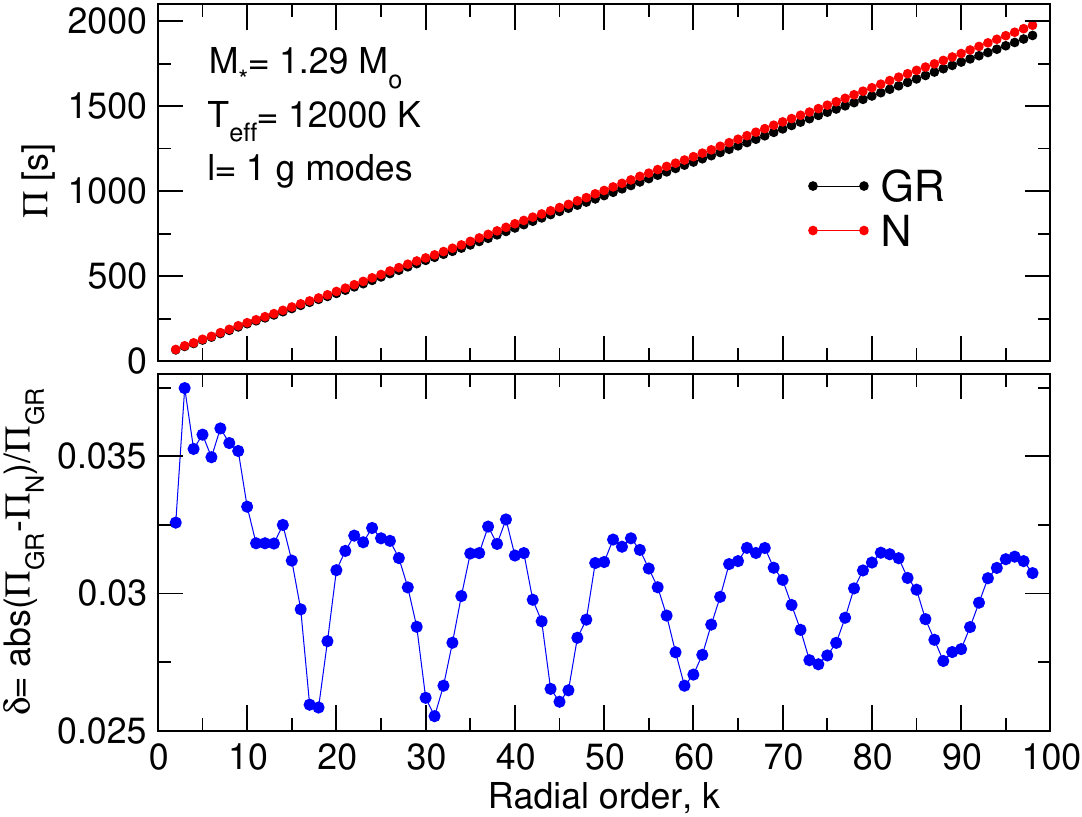}
        \caption{Upper panel: $\ell= 1$ $g$-mode periods in terms of the radial order ($k$) for a  WD model with $M_{\star}= 1.29 M_{\sun}$ and $T_{\rm eff}\sim 12\,000$ K for the relativistic (GR, black dots)  and Newtonian (N, red dots) cases. Lower panel: the absolute value of the relative difference between the periods of the GR and the N cases, $\delta= |\Pi_{\rm GR}-\Pi_{\rm N}|/\Pi_{\rm GR}$ (blue).} 
        \label{fig:period_vs_radial_order-1.29}
\end{figure}

\begin{figure}
\includegraphics[width=1.\columnwidth]{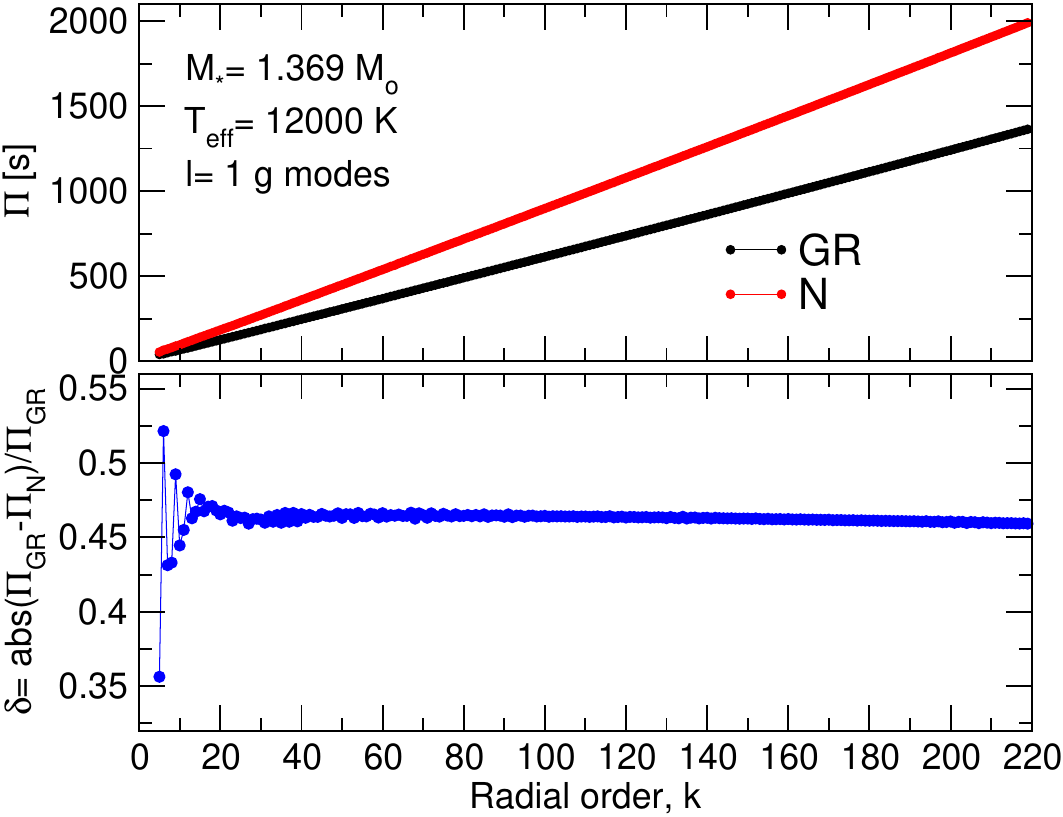}
        \caption{Same as Fig. \ref{fig:period_vs_radial_order-1.29}, but for the case of a WD 
        model with $M_{\star}= 1.369 M_{\sun}$.} 
        \label{fig:period_vs_radial_order-1.369}
\end{figure}

Since there are substantial differences in the separation of $g$-mode periods in the GR and N cases, it is natural to expect significant
differences in the individual pulsation periods ($\Pi$). In the upper panels of Figs. \ref{fig:period_vs_radial_order-1.29} and \ref{fig:period_vs_radial_order-1.369}, we compare the periods of the GR and N cases for the less massive ($1.29 M_{\sun}$) and the most massive ($1.369 M_{\sun}$) WD model considered in this work ($T_{\rm eff} \sim 12\,000$ K). $M_{\star}= 1.369 M_{\sun}$ corresponds to the maximum possible value in the calculations of \cite{2022A&A...668A..58A}, above which the models become unstable with respect to GR effects. 
It is clear from these figures that the periods in the relativistic case are shorter than those in the Newtonian case, with the absolute differences becoming larger with increasing $k$. This is mainly due to the structural differences of the equilibrium models in the GR case in relation to the N case (smaller radii and larger gravities characterising the relativistic WD models, see Fig. \ref{fig:radio-gravedad}), and to a much lesser extent, due to the differences in the relativistic treatment of the pulsations in comparison with the Newtonian one.

To quantify the impact of GR on the period spectrum, we have plotted 
in the lower panel of each figure the absolute value of the relative differences between the GR periods and the N periods, $\delta= |\Pi_{\rm GR}-\Pi_{\rm N}|/\Pi_{\rm GR}$,  versus the radial order. These differences are smaller than $\sim 0.035$ for the less massive model ($1.29~M_{\sun}$, Fig. \ref{fig:period_vs_radial_order-1.29}), but they become as large as $\sim 0.5$ for the most massive models ($1.369~M_{\sun}$, Fig. \ref{fig:period_vs_radial_order-1.369}). We conclude that, for ultra-massive WDs with masses in the range $1.29 \leq M_{\star}/M_{\sun} \leq 1.369$, the impact of GR on the pulsations is important, resulting in changes from  $\sim 4 \%$ to $\sim 50 \%$ in the values of $g$-mode periods. 

\begin{figure}
\includegraphics[width=1.\columnwidth]{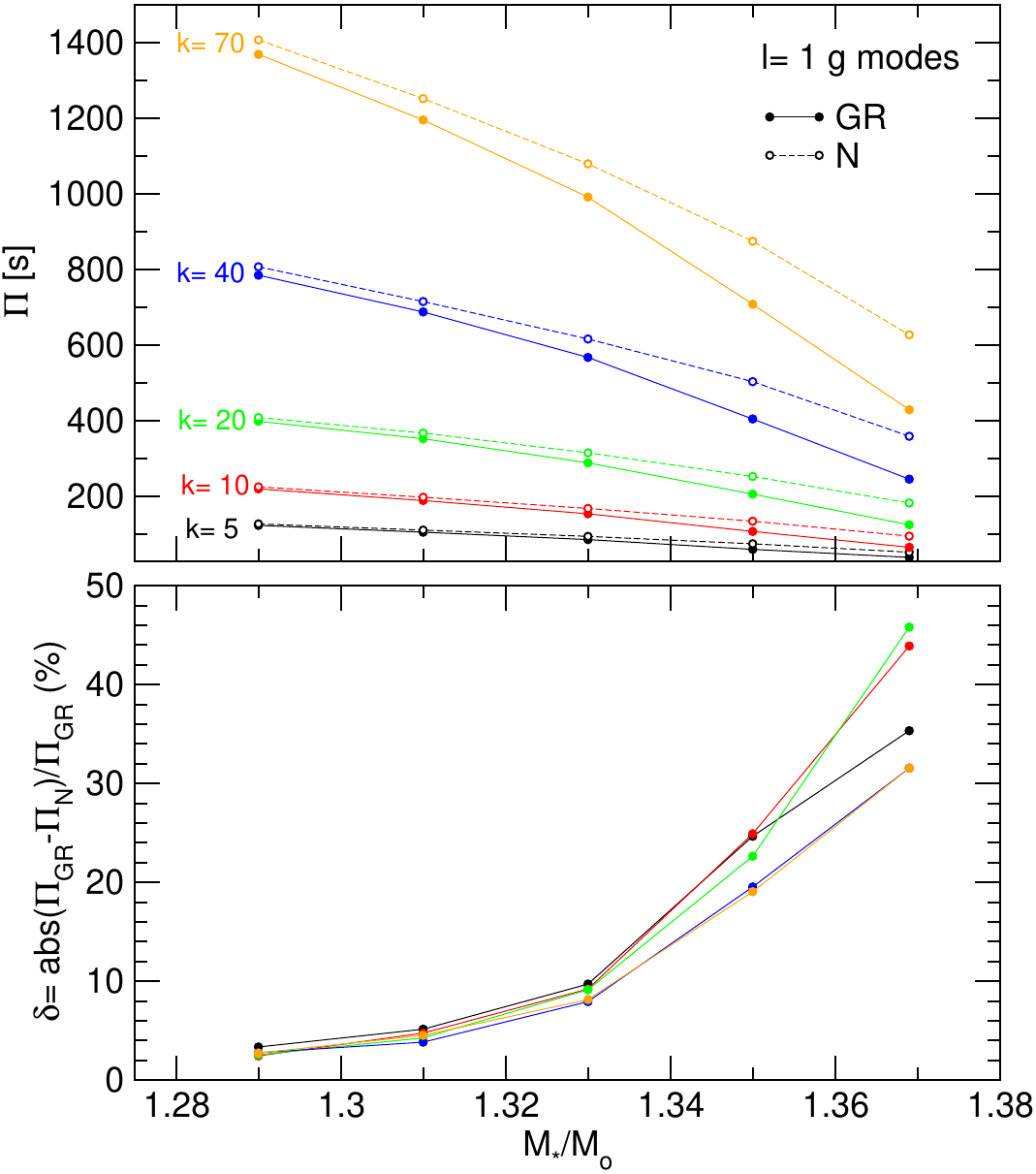}
        \caption{Upper panel: the periods of selected $\ell= 1$ $g$ modes ($k= 5, 10, 20, 40$ and 70) in terms of the stellar mass, for the GR case (solid lines with filled dots) and the N case (dashed lines with hollow dots). Lower panel: the absolute relative difference $\delta$ (percentage \%) between the relativistic and Newtonian periods, as a function of the stellar mass.} 
        \label{fig:deltaP_mass}
\end{figure}

Another way to visualise the impact of GR on the pulsation periods is to plot the periods for the GR and N cases in terms of stellar mass. 
We display in the upper panel of Fig. \ref{fig:deltaP_mass} the periods of selected $g$ modes (with radial order $k= 5, 10, 20, 40$ and 70) in terms of the stellar mass for the GR and the N cases. In the lower panel, we show the absolute value of the relative difference $\delta$ (percentage \%) between the relativistic and Newtonian periods, as a function of the stellar mass. The relative differences in the periods exhibit an exponential growth with stellar mass, without appreciable dependence on the radial order (see also Figs. \ref{fig:period_vs_radial_order-1.29} and \ref{fig:period_vs_radial_order-1.369}). The behaviour of $\delta$ with the stellar mass visibly mirrors the exponential increase in the relative differences between the relativistic and Newtonian stellar radii and surface gravities, as seen  in Fig. \ref{fig:radio_masa}. 

At first glance, the relative differences $\delta$  might seem larger than expected, given recent work on periods in average-mass WDs by \citet{2023arXiv230413055R}. For the simple models considered there, it was shown that $\delta \sim z \sim 10^{-4}$ for a WD with $M_{\star}\approx0.6M_{\odot}$.  However, considering their Figure 4, it is possible for stars with high central concentrations, such as ultra-massive WDs, that $\delta$ can be larger than $z$, consistent with our present findings.  To confirm this,
we also carried out pulsational calculations on a simplified stratified Chandrasekhar-type equilibrium model  that mimics a $\sim 1.3~M_{\sun}$ ultra-massive WD, in the case of Newtonian gravity and in the Post-Newtonian approximation, following the process in \citet{2023arXiv230413055R}. These calculations and their results are presented in the Appendix \ref{Appendix2}. The comparison of the periods in both cases indicates a relative difference of the order of $10^{-2}$, in complete agreement with the results obtained here for our WDs models of $1.29 M_{\sun}$ and $1.31 M_{\sun}$ (see Figs. \ref{fig:period_vs_radial_order-1.29} and \ref{fig:deltaP_mass}).

\begin{figure}
\includegraphics[width=1.\columnwidth]{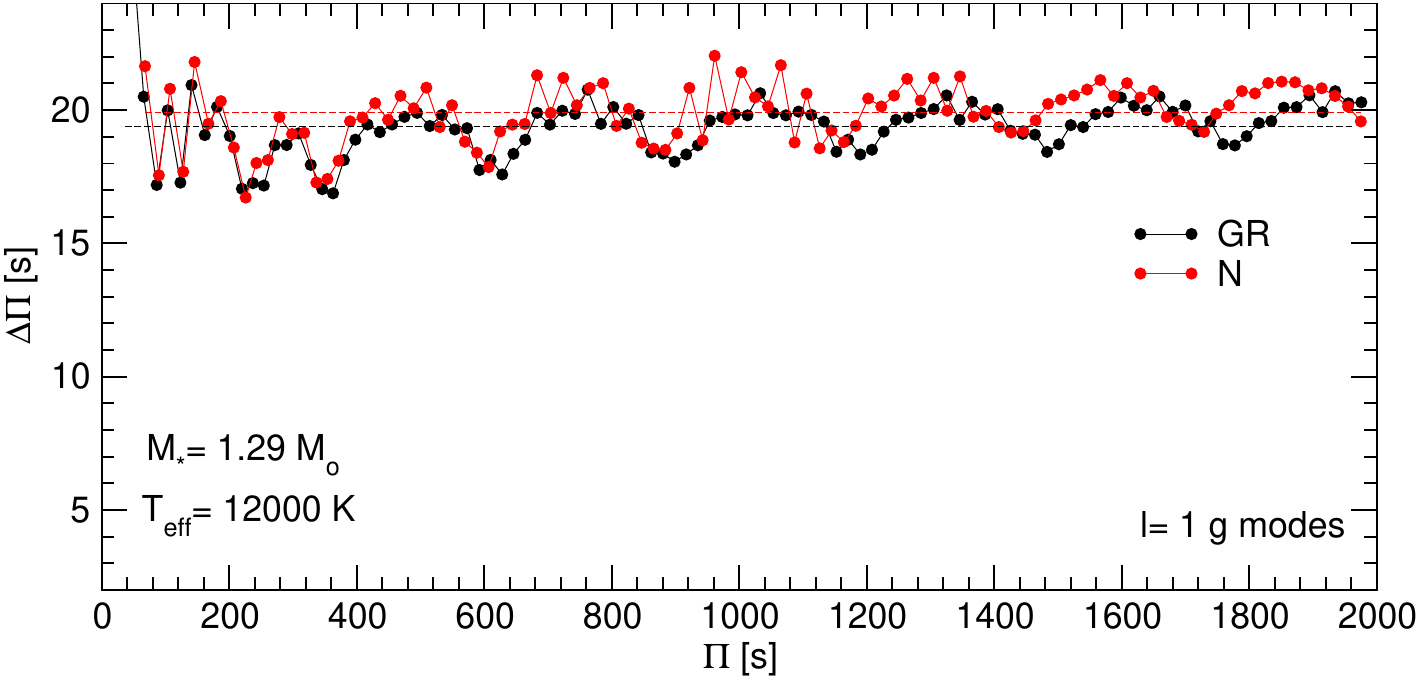}
        \caption{Dipole ($\ell= 1$) forward period spacing ($\Delta \Pi= \Pi_{k+1}-\Pi_{k}$) versus periods, corresponding to a  WD model with $M_{\star}= 1.29 M_{\sun}$ and $T_{\rm eff} \sim 12\,000$ K for the relativistic case (red),  and the Newtonian case (black). The horizontal dashed lines correspond to the asymptotic period spacings for both cases.} 
        \label{fig:del_period_vs_period-1.29}
\end{figure}

\begin{figure}
\includegraphics[width=1.\columnwidth]{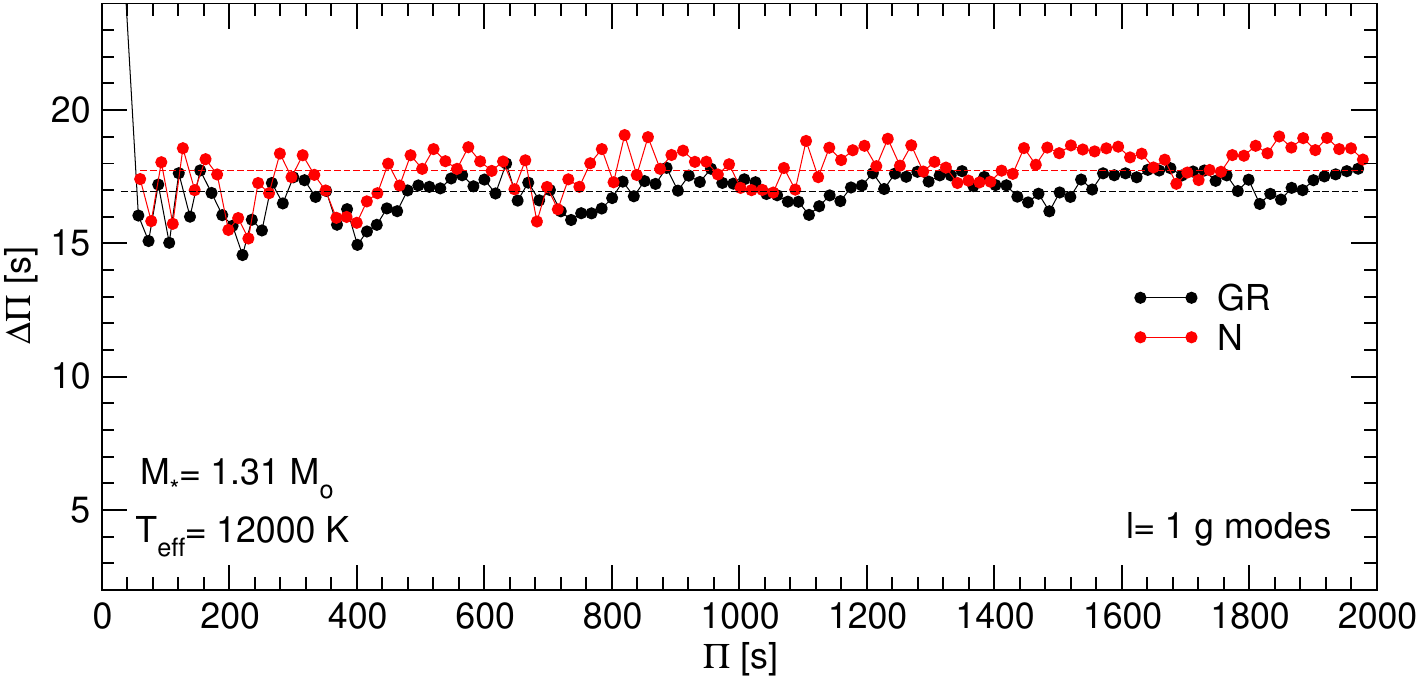}
        \caption{Same as Fig. \ref{fig:del_period_vs_period-1.29}, but for a  WD model with $M_{\star}= 1.31 M_{\sun}$.} 
        \label{fig:del_period_vs_period-1.31}
\end{figure}

\begin{figure}
\includegraphics[width=1.\columnwidth]{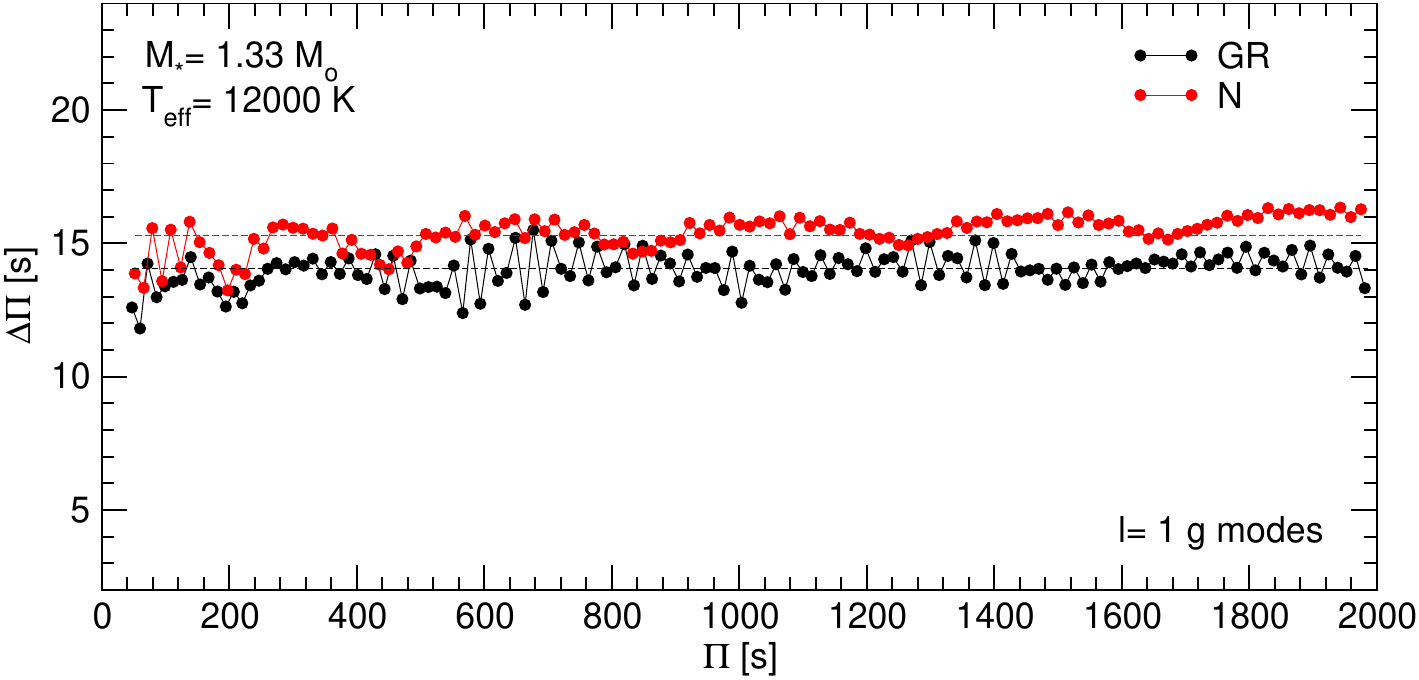}
        \caption{Same as Fig. \ref{fig:del_period_vs_period-1.29}, but for a  WD model with $M_{\star}= 1.33 M_{\sun}$.} 
        \label{fig:del_period_vs_period-1.33}
\end{figure}

\begin{figure}
\includegraphics[width=1.\columnwidth]{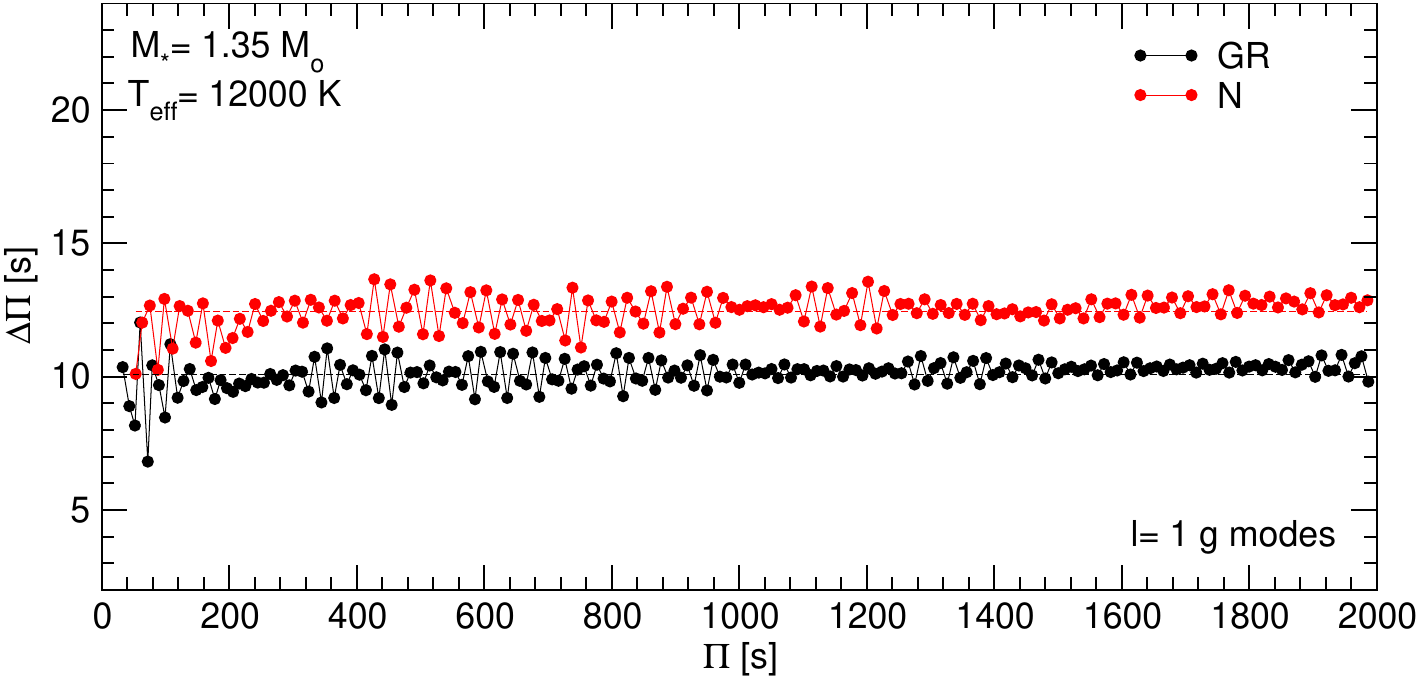}
        \caption{Same as Fig. \ref{fig:del_period_vs_period-1.29}, but for a  WD model with $M_{\star}= 1.35 M_{\sun}$.} 
        \label{fig:del_period_vs_period-1.35}
\end{figure}

\begin{figure}
\includegraphics[width=1.\columnwidth]{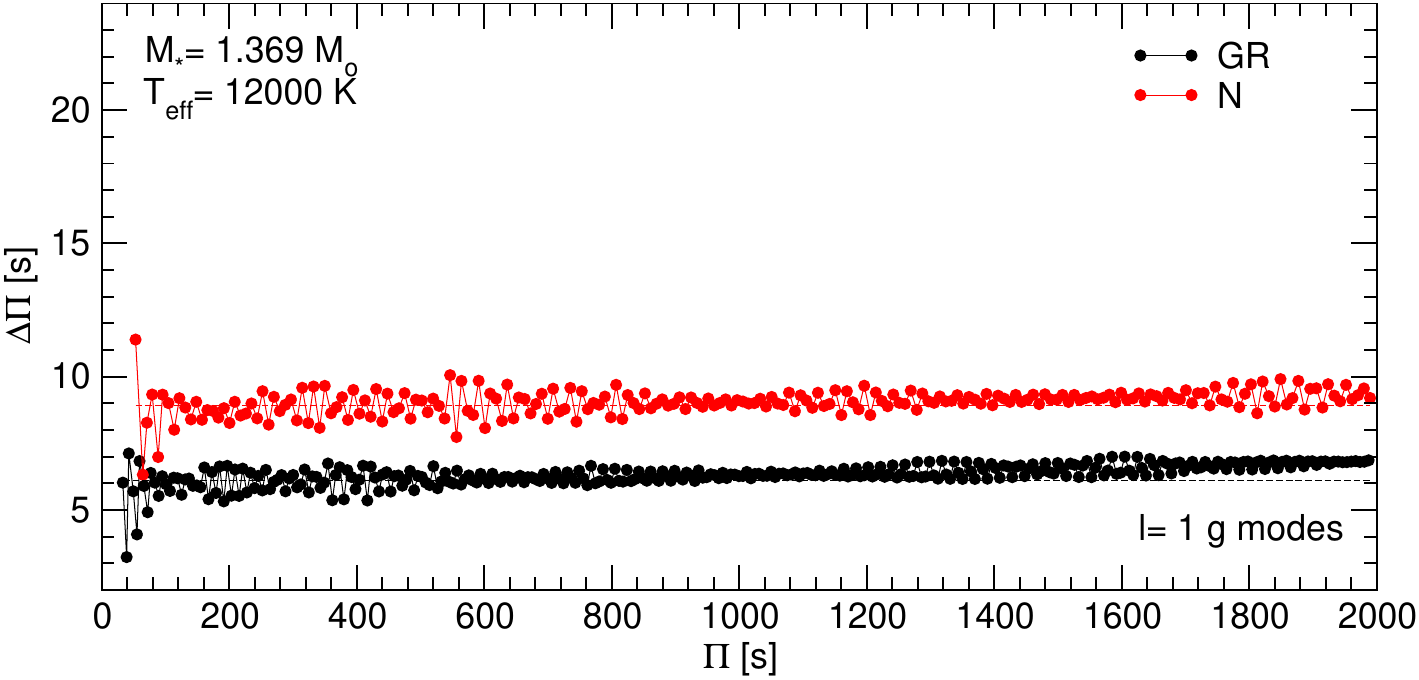}
        \caption{Same as Fig. \ref{fig:del_period_vs_period-1.29}, but for a  WD model with $M_{\star}= 1.369 M_{\sun}$.} 
        \label{fig:del_period_vs_period-1.369}
\end{figure}

It is interesting to examine how the period spacings versus periods change depending on whether we consider the GR case or the N case. We define the forward period spacing as $\Delta \Pi= \Pi_{k+1}-\Pi_k$. The dipole ($\ell= 1$) forward period spacing in terms of the periods is plotted in Figs. \ref{fig:del_period_vs_period-1.29} to \ref{fig:del_period_vs_period-1.369} for WD models with stellar masses between 
1.29 and $1.369~M_{\odot}$ and $T_{\rm eff}= 12\,000$ K. 
We have adopted the same range in the $y-$axis in order to make the 
comparison of the results between the different stellar masses clearer.
These figures show that, in general, the period spacing is larger in the N case than in the GR case, and that this difference becomes larger as the stellar mass increases. This is expected based on the behaviour of the asymptotic period spacing (see Fig.\ref{fig:asymptotic}), which is indicated with horizontal dashed lines.

\subsection{The case of the ultra-massive ZZ Ceti star WD~J0049$-$2525}
\label{WDJ004917}

The ultra-massive DA WD star WD~J004917.14$-$252556.81 ($T_{\rm eff}= 13\,020\pm460$~K, $\log g= 9.341\pm0.036$) is the most massive pulsating WD known to date \citep[][]{2023MNRAS.522.2181K}. It shows only two periods, at $\sim 209$~s and $\sim 221$~s, which are insufficient to find a single seismological model that would give us details of its internal structure.  Extensive follow-up time-series photometry could allow discoveries of a significant number of additional pulsation periods that would help to probe its interior. Considering the ONe-core WD evolutionary models of \cite{2022A&A...668A..58A}, WD~J0049$-$2525 has $M_{\star}= 1.283 \pm 0.008~M_{\sun}$ in the Newtonian gravity, or $M_{\star}= 1.279\pm 0.007~M_{\sun}$ if we adopt the GR treatment.  This heavyweight ZZ Ceti, in principle, could be considered as an ideal target to explore the relativistic effects on ultra-massive WD pulsations. However, the difference between the relativistic and Newtonian mass of this target is tiny. A difference of only $0.004 M_{\sun}$ is even smaller than the uncertainties in the mass estimates. This
small difference is due to the star being just slightly below the lower mass limit for the relativistic effects to be important\footnote{That is, $M_{\star} \sim 1.3 M_{\sun}$, the lower limit of the mass regime of the so-called {\it "relativistic ultra-massive WDs} \citep{2023MNRAS.523.4492A}.}. 

Fig. \ref{fig:deltaP_mass} (see also Fig. \ref{fig:del_period_vs_period-1.29}) demonstrates that the effects of the GR on the $g$-mode periods of WD~J0049$-$2525 are less than $\sim 1 \%$. Although extremely important for being the most massive pulsating WD star known, WD~J0049$-$2525 is not massive enough for the exploration of the GR effects on WD pulsations. We conclude that, to be able to study the effects of GR on WD pulsations, we have to wait for the discovery and monitoring of even more massive pulsating WDs, especially the ones with $M_{\star}\gtrsim 1.33~M_{\sun}$.

\subsection{Prospects for finding pulsating WDs where GR effects are significant}
\label{future}

\begin{figure}
\centering
\includegraphics[width=3.4in, clip=true, trim=0.3in 2in 0.3in 1.4in]{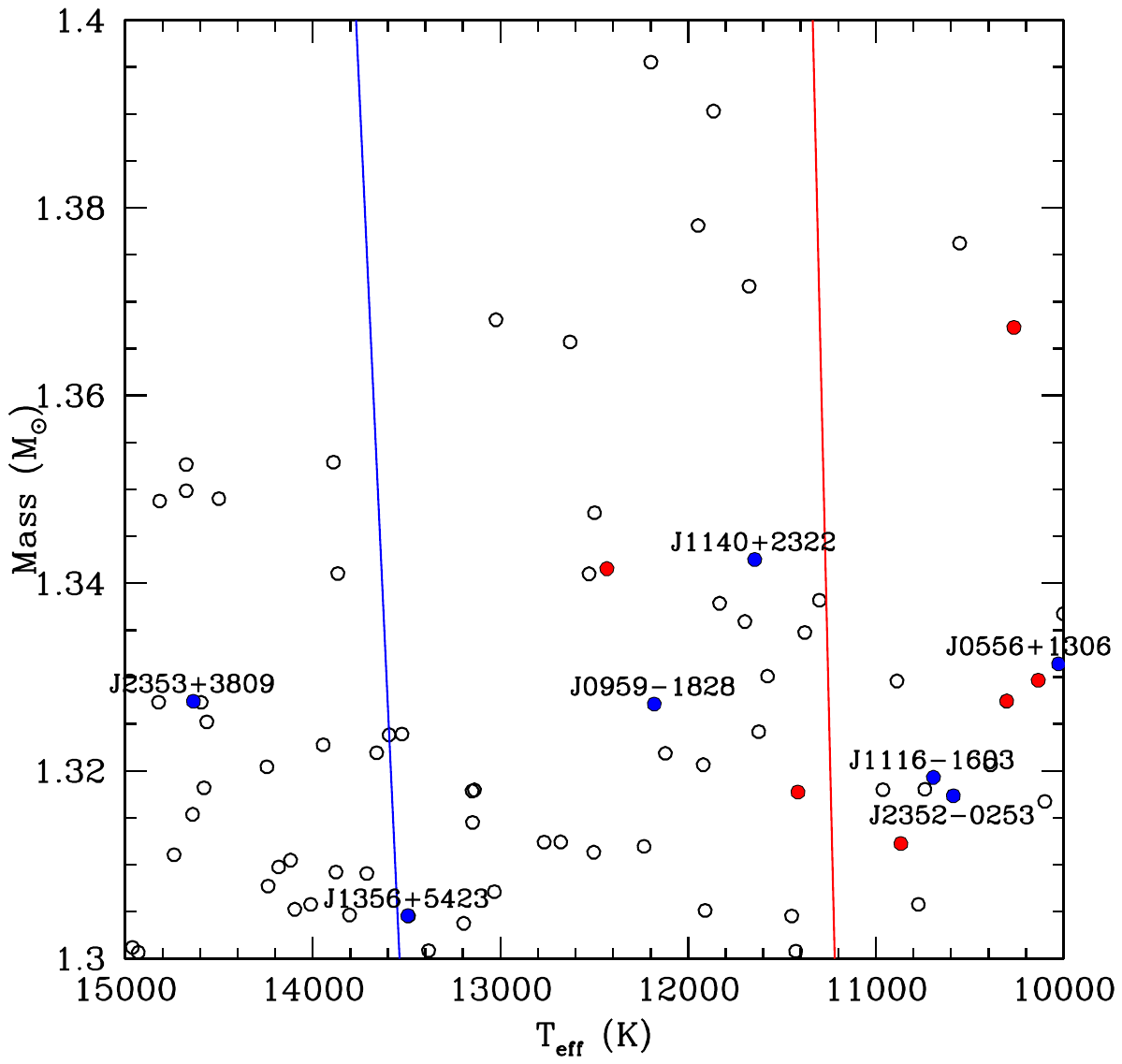}
\caption{Masses and effective temperatures for high probability WD candidates with $M_{\star} \geq 1.3~M_{\odot}$ in the {\it Gaia} EDR3 WD sample from \citet{2021MNRAS.508.3877G} assuming CO cores.  For ONe cores, the masses would be lower on
average by $0.04-0.05~M_{\odot}$. The blue and red lines show the empirical boundaries of the ZZ Ceti instability strip from \citet{2020AJ....160..252V} extrapolated to higher masses. Blue and red dots show the spectroscopically confirmed DA and DC/magnetic WDs, respectively. }
\label{figstrip}
\end{figure}

Figure \ref{figstrip} shows the masses and effective temperatures for high probability ($P_{\rm WD}\geq0.9$) WD candidates with
$M_{\star}\geq1.3~M_{\odot}$ in the Gaia EDR3 WD sample from \citet{2021MNRAS.508.3877G} assuming CO cores. Here we limit the sample to the temperature
range near the ZZ Ceti instability strip. The blue and red lines show the boundaries of the instability strip from \citet{2020AJ....160..252V} extrapolated to higher masses. There are 78 objects in this sample, including 7 spectroscopically confirmed DA WDs (labelled in the figure)
and 6 magnetic or DC WDs. \citet{2023MNRAS.518.2341K} found that only 48\% of the $M_{\star}\approx 1.3\, {\rm M}_{\odot }$ WDs within 100 pc are DA WDs, with the rest being strongly magnetic (40\% of the sample) or WDs with unusual atmospheric compositions (hot DQ, DBA, DC etc). Hence, follow-up spectroscopy is required to identify the DA WDs in this sample. 

\citet{2023MNRAS.518.2341K} presented time-series photometry for the five DA WDs cooler than $13\,000$ K in Fig. \ref{figstrip}. They did not detect
any significant variations in four of the targets, and their observations were inconclusive for J0959$-$1828. Nevertheless, there are a number of relativistic ultra-massive WD candidates that may fall within the ZZ Ceti instability strip, and therefore may exhibit pulsations. The masses shown here are based on the CO-core evolutionary models, for ONe cores the masses would be lower on average by $0.04-0.05~M_{\odot}$. Even then, there are 9 candidates with $M_{\star} > 1.35~M_{\odot}$ and up to $1.39~M_{\odot}$ (assuming a CO core) near the instability strip. If confirmed, such targets would be prime examples of objects where GR effects would have a significant impact on their pulsation properties.

Unfortunately, the observational errors in temperatures and masses of these targets based on {\it Gaia} photometry and parallaxes \citep{2021MNRAS.508.3877G} are too large to effectively identify the best targets for follow-up. For example, 64 of the 78 objects shown here have temperature errors larger than 2000 K, roughly the width of the instability strip, and 62 have errors in mass that are larger than $0.1~M_{\odot}$. Hence, further progress on understanding the GR effects on WD pulsation will require spectroscopic and time-series observations of a relatively large sample of candidates to identify genuine pulsating ultra-massive WDs with $M_{\star}\gtrsim 1.33~M_{\sun}$.  In addition, the median $G-$band magnitude for these 78 objects is 20.25 mag. Hence, 4-8m class telescopes would be needed to confirm pulsating DA WDs in this sample.

\section{Summary and conclusions}
\label{conclusions}

In this paper, we have assessed for the first time the impact of GR on the  $g$-mode period spectra of ultra-massive ZZ Ceti stars. To this end, we pulsationally analysed fully evolutionary ONe-core ultra-massive WD models with masses from $1.29$ to $1.369 M_{\sun}$ computed in the frame of GR \citep{2022A&A...668A..58A}. We employed the {\tt LPCODE} and {\tt LP-PUL} evolutionary and pulsation codes, respectively, adapted for relativistic calculations. In particular, for the pulsation analysis, we  considered the relativistic Cowling approximation. Our study  is consistent with \cite{2023arXiv230413055R}, considering the high central compactness of the stars studied here.  The study of pulsating ultra-massive WDs in the context of GR is timely considering the increasing rate of discovery of very high-mass objects \citep[e.g.,][]{2020ApJ...898...84K, 2021MNRAS.503.5397K, 2020NatAs...4..663H, 2021Natur.595...39C, 2022MNRAS.511.5462T,  2023MNRAS.518.2341K}, the discovery of the ZZ Ceti WD~J004917.14$-$252556.81 \citep[the most massive pulsating WD currently known,][]{2023MNRAS.522.2181K}, and the possibility of finding even more massive pulsating objects in the near future. This is particularly relevant in view of the space-based surveys like {\it TESS} and {\it ULTRASAT} and wide-field ground-based surveys like the LSST and BlackGEM.

We find that the Brunt-V\"ais\"al\"a and Lamb frequencies are larger for the Relativistic case compared to the Newtonian case, as a result of relativistic models having smaller radii and higher gravities. This has the important consequence that the typical separation between consecutive $g$-mode periods is smaller in the relativistic case than in the Newtonian computations, with percentage differences of up to 48\% in the case of the most massive model ($1.369~M_{\sun}$). We assessed the dipole period spectrum of $g$ modes of our ultra-massive WD models for the Newtonian and the relativistic cases, and found that the periods in the GR case are shorter than in the Newtonian computations. In particular, for the less massive model ($1.29~M_{\sun}$), these relative differences are smaller than $\sim 0.035$, but the variations reach values as large as $\sim 0.5$ for the most massive model ($1.369~M_{\sun}$). 

We conclude that, for ultra-massive DA WDs models with masses in the range that we have considered in this paper ($1.29 \leq M_{\star}/M_{\sun} \leq 1.369$) and effective temperatures typical of the ZZ Ceti instability strip, GR does matter in computing the adiabatic $g$-mode pulsations, resulting in periods that are between $\sim 4$ and $\sim 50\%$ shorter, depending on the stellar mass, when a relativistic treatment is adopted instead of a Newtonian one. 
This suggests that the effects of GR on the structure and pulsations of WDs with masses $\gtrsim 1.29 M_{\sun}$ cannot be ignored in asteroseismological analysis of ultra-massive ZZ Ceti stars and likely other classes of pulsating WDs.

\section*{Acknowledgements}

 We  wish  to  thank  the  suggestions  and
comments of an anonymous referee that improved the original
version of this work. Part of this work was supported by AGENCIA through the Programa de Modernizaci\'on Tecnol\'ogica BID 1728/OC-AR, by the
PIP 112-200801-00940 grant from CONICET, by the National Science Foundation under grants AST-2205736, PHY-2110335, the National Aeronautics
and Space Administration under grant 80NSSC22K0479, and the US DOE under contract  DE-AC05-00OR22725. ST acknowledges support from MINECO under the PID2020-117252GB-I00 grant and by the AGAUR/Generalitat de Catalunya grant SGR-386/2021. MC acknowledges grant RYC2021-032721-I, funded by MCIN/AEI/10.13039/
501100011033 and by the European Union NextGenerationEU/PRTR.
This  research has  made use of  NASA Astrophysics Data System. 

\section*{Data Availability Statement}

The data underlying this article are available upon request.

%



\bibliographystyle{mnras}
\bibliography{bibliografia.bib}


\appendix
\section{Relativistic expression for the Brunt-V\"ais\"al\"a frequency 
in the ``modified Ledoux'' prescription}
\label{Appendix1}

We start from the relativistic expression for the Brunt-V\"ais\"al\"a frequency, obtained according to its definition (Eq.~\ref{eq:bv_def}).  This expression can be derived by considering slight buoyant perturbations of a fluid packet within the stellar medium, as detailed in \cite{reese..bostonphd}. In the GR case, $N^2$ is given by:

\begin{equation}
  N^2= -\frac{c^2}{2} \nu^{\prime} e^{-\lambda}
  \left[\frac{1}{\rho+P/c^2} \left(\frac{d\rho}{dr}\right)- 
  \frac{1}{\Gamma_1 P} \left(\frac{dP}{dr}\right)\right],   
  \label{ap1}  
\end{equation} 

\noindent which reduces to the Newtonian result in the limit $c\rightarrow\infty$. 
In what follows, we will derive expressions for the first and second members 
inside the brackets of Eq.~(\ref{ap1}).
In the Newtonian case, if stellar plasma is composed by $M$ atomic species with fractional abundances $X_i$, the equation of state can be written as:

\begin{equation}
P= P(\rho,T,\{X_i\}),   
\label{ap2}
\end{equation}

\noindent where $i=1, \cdots, M-1$, and $\sum_{i=1}^{M-1} X_i+X_M= 1$.  In the relativistic case, 
we have, instead:

\begin{equation}
P= P(n,T,\{X_i\}),   
\label{ap3}
\end{equation}

\noindent where $n$ is the baryonic number density. Following \cite{1991ApJ...367..601B}, we can differentiate $P$ from Eq.~(\ref{ap3}) and write:

\begin{equation}
d\ln P= \chi_n d\ln n + \chi_T d\ln T + \sum_{i= 1}^{M-1} \chi_{X_i} d\ln X_i,
\label{ap4}
\end{equation}

\noindent where $\chi_{\rm T}$, $\chi_{n}$, and $\chi_{X_i}$ are given by 
Eqs.~(\ref{chi_T_chi_n_chi_Xi}). Now, from Eq. (\ref{ap4}) 
we have:

\begin{equation}
\frac{d\ln n}{d\ln P}= \frac{1}{\chi_n} - \frac{\chi_T}{\chi_n} \nabla - \frac{1}{\chi_n} \sum_{i= 1}^{M-1} \chi_{X_i} \frac{d\ln X_i}{d\ln P},
\label{ap5}
\end{equation}

The relativistic adiabatic exponent, $\Gamma_1$, is defined as:

\begin{equation}
\Gamma_1=\left(\frac{\partial \ln P}{\partial \ln n}\right)_{\rm ad}
\label{ap6}
\end{equation}

\noindent Following \citet[][Eqs.~6.6 and 13.24]{1990sse..book.....K}, we can write 

\begin{equation}
\Gamma_1= \frac{1}{\alpha-\delta \nabla},
\label{ap7}
\end{equation}

\noindent where:

\begin{equation}
\alpha= \left(\frac{d\ln \rho}{d\ln P}\right)_T,\ 
 \ \delta= -\left(\frac{d\ln \rho}{d\ln T}\right)_P.
\label{ap8}
\end{equation}

\noindent From the definition of $\chi_{T}$ and  $\chi_{n}$ 
(Eqs.~\ref{chi_T_chi_n_chi_Xi}), and using the property of the partial derivatives $(\partial f/\partial y)_{x}= 
- (\partial f/\partial x)_{y} \cdot (\partial x/\partial y)_{f}$, we have:

\begin{equation}
\alpha= \frac{1}{\chi_n},\ 
\  \ \delta= \frac{\chi_T}{\chi_n}.
\label{ap9}
\end{equation}

\noindent so that $\Gamma_1$ can be written as:

\begin{equation}
\frac{1}{\Gamma_1}= \frac{\left(1-\chi_T \nabla_{\rm ad}\right)}{\chi_n}.
\label{ap10}
\end{equation}

The first law of thermodynamics in GR can be written \citep[Eq. 2.12 of][converted to standard, non-geometrized units]{1967hea3.conf..259T} 
as

\begin{equation}
d\rho= \left[\frac{\rho+ P/c^2}{n}\right] dn + \frac{T}{c^2} n\ ds + \sum_k 
\mu_k n\ dX_k, 
\label{ap12}
\end{equation}

\noindent where $T$, $s$, and  $\mu_k$ are the temperature, the entropy per
baryon, and the nuclear chemical potential of the species $k$, respectively. 

\noindent If we now assume isentropic changes ($ds= 0$) and suppose that the abundances of the nuclear species do not change ($dX_k= 0$) \citep[Eq. 2.14 of ][]{1967hea3.conf..259T}, 
then by differentiating with respect to $r$, we finally have:

\begin{equation}
\frac{d\rho}{dr}= \left[\frac{\rho+ P/c^2}{n}\right] \frac{dn}{dr}. 
\label{ap13}
\end{equation}

\noindent Eq.~(\ref{ap13}) can be written as:

\begin{equation}
\frac{1}{\rho+P/c^2} \left(\frac{d\rho}{dr}\right)= \frac{1}{n}\frac{d n}{dr}
\label{ap14}
\end{equation}

\noindent Substituting Eq. (\ref{ap14}) in Eq.~(\ref{ap1}), using the static TOV equation of GR \citep{1939PhRv...55..364T, 1939PhRv...55..374O},

\begin{equation}
\frac{dP}{dr}= -\frac{1}{2} \left(\rho+\frac{P}{c^2} \right) c^2 \nu^{\prime}, 
\label{ap15}
\end{equation}

\noindent and employing Eqs.~(\ref{ap5}) and (\ref{ap10}), we obtain:

\begin{equation}
  N^2= \left( \frac{1}{2} c^2 \nu^{\prime}\right)^2 e^{-\lambda} 
\left(\frac{\rho+P/c^2}{P} \right) 
\frac{\chi_T}{\chi_n} \left[\nabla_{\rm ad} - \nabla -  
\frac{1}{\chi_T} \sum_{i= 1}^{M-1} \chi_{X_i} \frac{d\ln Xi}{d\ln P} \right],
\label{ap16}  
\end{equation} 

\noindent where the last term inside the brackets is the Ledoux term $B$ (Eq.~\ref{B_Ledoux}). Thus,
we finally obtain:

\begin{equation}
  N^2= \left( \frac{1}{2} c^2 \nu^{\prime}\right)^2 e^{-\lambda} 
\left(\frac{\rho+P/c^2}{P} \right) 
\frac{\chi_T}{\chi_n} \left[\nabla_{\rm ad} - \nabla + B \right],
\label{ap17}  
\end{equation} 

\section{Validation with a toy model based on Chandrasekhar's models}
\label{Appendix2}

As validation of the results presented in this paper, in particular the size of the relative difference in the periods, we have carried out pulsation calculations on a toy model based on Chandrasekhar's models, with a stellar mass $M_{\star} \sim  1.3M_{\odot}$. This model has a cold degenerate-electron equation of state featuring a near-surface chemical transition from $\mu_e= 2$ to $\mu_e= 1$, simulating a surface H layer. Thus, this simple model mimics the structure of a stratified 
realistic ultra-massive WD model. Following the Post-Newtonian method described in \citet{2023arXiv230413055R}, we have compared the fourth-order nonradial Newtonian pulsations to the nonradial GR pulsations for this toy model for several $g$, $f$, and $p$ modes with low radial orders $k$ for harmonic degree $\ell = 1, 2$ and 3. 
We show the results in Table \ref{tab:CHWD++}. The relative differences we obtain for $g$ modes are, on average, $\sim 2.65 \times 10^{-2}$ (column 4), which is 
consistent with the results shown in Figs. \ref{fig:period_vs_radial_order-1.29} and \ref{fig:deltaP_mass} for the cases of ultra-massive WD models with $M_{\star}= 1.29 M_{\sun}$ and $M_{\star}= 1.31 M_{\sun}$, respectively. 

\begin{table}
\caption{Periods for several low-order $p$, $f$, and $g$ modes corresponding to the Newtonian gravity computations (column 2) and the GR computations (column 3) for a stratified degenerate electron gas model, both with $M_{\star}= 1.29966 M_{\odot}$.  The Newtonian model uses $y_0=6.385$, with $z=6.6012\times10^{-4}$, while the  GR model uses $y_0=6.779$, with $z=6.3047\times10^{-4}$. Column 3 gives the relative difference, which is commensurate with those in Figure \ref{fig:period_vs_radial_order-1.29}. }
\label{tab:CHWD++}
\begin{center}
\begin{tabular}{crrcrrcc}
\hline
	&\multicolumn{1}{c}{Newtonian}
	&\multicolumn{1}{c}{Post-Newtonian}
	&  \\
mode $_{\ell,k}$
	& \multicolumn{1}{c}{$\Pi$ (s)}
	& \multicolumn{1}{c}{$\Pi$ (s)}
	& rel. diff. \\
\hline\hline
    $p_{1,1}$& 1.0495784& 0.9782581 & $1.52\times10^{-3}$\\
    $g_{1,1}$& 16.2675050& 14.7865834 & $2.66\times10^{-2}$\\
    $g_{1,2}$& 35.7515552& 32.4932485 & $2.67\times10^{-2}$\\
    $g_{1,3}$& 54.4264093& 49.4686549 & $2.67\times10^{-2}$\\
    $g_{1,4}$& 72.9174383& 66.2778687 & $2.66\times10^{-2}$\\
    $g_{1,5}$& 91.3355203& 83.0211550 & $2.66\times10^{-2}$\\
    $g_{1,6}$& 109.7169819& 99.7313589 & $2.66\times10^{-2}$\\
    $g_{1,7}$& 128.0773819& 116.4225439 & $2.66\times10^{-2}$\\
    $g_{1,8}$& 146.4245279& 133.1017651 & $2.65\times10^{-2}$\\
\noalign{\smallskip}
\hline
\noalign{\smallskip}
    $f_{2}$& 1.2909440& 1.1915768 & $1.12\times10^{-2}$\\
    $g_{2,1}$& 9.4164937& 8.5607991 & $2.64\times10^{-2}$\\
    $g_{2,2}$& 20.6825679& 18.8000960 & $2.66\times10^{-2}$\\
    $g_{2,3}$& 31.4771109& 28.6128454 & $2.66\times10^{-2}$\\
    $g_{2,4}$& 42.1642251& 38.3283552 & $2.65\times10^{-2}$\\
    $g_{2,5}$& 52.8083359& 48.0049281 & $2.65\times10^{-2}$\\
    $g_{2,6}$& 63.4306072& 57.6617194 & $2.65\times10^{-2}$\\
    $g_{2,7}$& 74.0401429& 67.3069695 & $2.65\times10^{-2}$\\
    $g_{2,8}$& 84.6415378& 76.9448380 & $2.65\times10^{-2}$\\
\noalign{\smallskip}
\hline
\noalign{\smallskip}
    $f_{3}$& 1.0796099& 1.0003529 & $7.39\times10^{-3}$\\
    $g_{3,1}$& 6.6828617& 6.0771126 & $2.62\times10^{-2}$\\
    $g_{3,2}$& 14.6647687& 13.3323883 & $2.64\times10^{-2}$\\
    $g_{3,3}$& 22.3087624& 20.2815701 & $2.64\times10^{-2}$\\
    $g_{3,4}$& 29.8754567& 27.1606785 & $2.64\times10^{-2}$\\
    $g_{3,5}$& 37.4108454& 34.0113796 & $2.64\times10^{-2}$\\
    $g_{3,6}$& 44.9300978& 40.8474210 & $2.64\times10^{-2}$\\
    $g_{3,7}$& 52.4397798& 47.6747566 & $2.64\times10^{-2}$\\
    $g_{3,8}$& 59.9432319& 54.4964175 & $2.64\times10^{-2}$\\
\hline\hline
\end{tabular}
\end{center}
\end{table}

\bsp	
\label{lastpage}
\end{document}